\documentclass[aps,prb,twocolumn,superscriptaddress,floatfix,10pt]{revtex4-2}
\usepackage{graphicx}
\usepackage{amsmath}
\usepackage{amssymb}
\usepackage{mathtools}
\usepackage{xcolor}
\usepackage{braket}
\usepackage{soul}
\usepackage[colorlinks=true, citecolor=blue, urlcolor=blue, linkcolor=blue,bookmarks=false,hypertexnames=true]{hyperref} 
\usepackage{appendix}
\usepackage{array}
\usepackage[column=O]{cellspace}
\newcolumntype{P}[1]{>{\centering\arraybackslash}p{#1}}
\allowdisplaybreaks
\usepackage{lipsum}
\usepackage{acro}
\usepackage{enumitem}
\usepackage{soul}

\DeclareAcronym{TMDC}{short=TMDC, long=transition metal dichalcogenide,}
\DeclareAcronym{DFT}{short=DFT, long=density functional theory,}
\DeclareAcronym{TB}{short=TB, long=tight-binding,}
\DeclareAcronym{QD}{short=QD, long=quantum dot,}
\DeclareAcronym{1D}{short=1D, long=one-dimensional,}
\DeclareAcronym{2D}{short=2D, long=two-dimensional,}
\DeclareAcronym{3D}{short=3D, long=three-dimensional,}
\DeclareAcronym{VB}{short=VB, long=valence band,}
\DeclareAcronym{CB}{short=CB, long=conduction band,}
\DeclareAcronym{UC}{short=UC, long=unit cell,}
\DeclareAcronym{PBE}{short=PBE, long=Perdew-Burke-Ernzerhof,}
\DeclareAcronym{PAW}{short=PAW, long=projector augmented wave,}
\DeclareAcronym{GGA}{short=GGA, long=generalized gradient approximation,}
\DeclareAcronym{SOC}{short=SOC, long=spin-orbit coupling,}
\DeclareAcronym{vdW}{short=vdW, long=van der Waals,}
\DeclareAcronym{BSE}{short=BSE, long=Bethe-Salpeter equation,}
\DeclareAcronym{hBN}{short=hBN, long=hexagonal boron nitride,}
\DeclareAcronym{DE}{short=DE, long=differential evolution,}
\DeclareAcronym{BZ}{short=BZ, long=Brillouin zone,}
\DeclareAcronym{NN}{short=NN, long=nearest neighbour}
\DeclareAcronym{NNN}{short=NNN, long=next-nearest neighbour}

\begin{document}
\graphicspath{}
\preprint{APS/123-QED}

\title{Electrically tunable MoSe$_2$/WSe$_2$ heterostructure-based quantum dot}
\date{\today}

\author{Katarzyna Sadecka}
\email{katarzyna.sadecka@pwr.edu.pl}
\affiliation{Institute of Theoretical Physics, Wroc\l aw University of Science and Technology, Wybrze\.ze Wyspia\'nskiego 27, 50-370 Wroc\l aw, Poland}
\affiliation{Department of Physics, University of Ottawa,
Ottawa, K1N6N5, Canada}

\author{Maciej Bieniek}
\affiliation{Institute of Theoretical Physics, Wroc\l aw University of Science and Technology, Wybrze\.ze Wyspia\'nskiego 27, 50-370 Wroc\l aw, Poland}

\author{Paulo E. {Faria~Junior}}
\affiliation{Institute of Theoretical Physics, University of Regensburg, 93040 Regensburg, Germany}

\author{Arkadiusz W\'ojs}
\affiliation{Institute of Theoretical Physics, Wroc\l aw University of Science and Technology, Wybrze\.ze Wyspia\'nskiego 27, 50-370 Wroc\l aw, Poland}

\author{Pawe\l{} Hawrylak}
\affiliation{Department of Physics, University of Ottawa,
Ottawa, K1N6N5, Canada}

\author{Jaros\l{}aw Paw\l{}owski}
\email{jaroslaw.pawlowski@pwr.edu.pl}
\affiliation{Institute of Theoretical Physics, Wroc\l aw University of Science and Technology, Wybrze\.ze Wyspia\'nskiego 27, 50-370 Wroc\l aw, Poland}

%%%%%%%%%%%%%%%%%%%%%%%%%%%%%%%%%%%%%%%%% Abstract %%%%%%%%%%%%%%%%%%%%%%%%%%%%%%%%%%%%%%%%%
\begin{abstract}

We describe here a theory of a quantum dot in an electrically tunable MoSe$_2$/WSe$_2$ heterostructure. Van der Waals heterostructures allow for tuning their electronic properties beyond their monolayer counterparts. We start by determining their electronic structure using density functional theory. We obtain the type-II band alignment and close in energy conduction band minima (valleys) at the $K$ and $Q$ points in the Brillouin zone. The valence band maxima, also energetically close, are located at the $K$ and $\Gamma$ points. By analyzing the Kohn-Sham wavefunctions, we describe the layer, spin, and orbital contributions. Next, we construct an \textit{ab initio}-based tight-binding model, which helps us to better understand the complexity of the interlayer interactions. We determine the effect of a vertical electric field, showing that vertical gating enables control of valleys extrema and their occupancy. Finally, we employ the tight-binding model to investigate laterally gated quantum dots and analyze the influence of a perpendicular electric field on their energy spectrum. Our results demonstrate that tuning the electric field enables control over the valley character of the quantum dot states, selectively localizing them in either the $K$ or $Q$ valleys, as evidenced by their characteristic degeneracies and wavefunctions.

\vspace{3mm}
\end{abstract}
\maketitle
%%%%%%%%%%%%%%%%%%%%%%%%%%%%%%%%%%%%%%%%%%%%%%%%%%%%%%%%%%%%%%%%%%%%%%%%%%%%%%%%%%%%%%%%%%

%%%%%%%%%%%%%%%%%%%%%%%%%%%%%%%%%%%%%% Introduction %%%%%%%%%%%%%%%%%%%%%%%%%%%%%%%%%%%%%%

%%%%%%%%%%%%%%%%%%%%%%%%%%%%%%%%%%%%%%
\section{Introduction}
\label{section:introduction}
%%%%%%%%%%%%%%%%%%%%%%%%%%%%%%%%%%%%%%

Single layers of \acp{TMDC} are \ac{2D} semiconductors enabling the exploration of many physical phenomena \cite{Splendiani_Galli_2010,Mak_Heinz_2010,Kadantsev_Hawrylak_2012,Wang_Urbaszek_2018,Liu2019Review}. For example, low-energy massive Dirac fermions possess a spin-valley pseudospin structure \cite{xiao2012coupled,bieniek2020band,bieniek2018zeeman,bieniek2022nanomaterials,Pawlowski2018ValleyQubit,Pawlowski_2019,Szulakowska2019magnetoX,Scrace_Hawrylak_2015,Srivastava_Imamoglu_2015, Zhou_Xiao_2015} that can be controlled using electromagnetic fields and accessed using polarized light. \acp{TMDC} are also the basic ''blocks'' for the construction of \ac{vdW} heterostructures \cite{geim2013van, Avalos-Ovando_2019}, in which many aspects of a monolayer controlability are preserved, in contrast to homobilayers. These heterostructures can be superior in some applications due to better control of electronic properties via electric field, light emission from interlayer excitons closer to telecommunication band, and enhanced valley coherence time~\cite{gao2017interlayer,calman2018indirect,jauregui2019electrical,Li2023quadrupolar}. Longer exciton lifetime opens new avenues for engineering long-range exciton transport and strongly interacting artificial bosonic simulators \cite{FowlerGerace_Butov_2024}. 

Quantum information platforms require initialization and control over individual quantum states. Electrical confinement and manipulation of charge carriers are essential factors. Circularly polarized light can populate (initialize) a given valley~\cite{Zeng2012valleys,Xiao2013TBmono}, nevertheless optical coupling of valley degrees of freedom in \acp{TMDC} can be realized only indirectly. However, valley coupling can be achieved electrically in \ac{TMDC}-based \acp{QD} \cite{Song2015WS2QD,song2015gate,lee2016coulomb,Wang2018TMDCQDs,pisoni2018gate,Davari_2020,Liu2019Review,Menaf2021Qubits,Pawlowski2021Qubit,Saleem2023Exciton,Korkusinski2023BLGQD,Sadecka2024Trion}. Different methods of realizing the electrical control over the valley degree of freedom have been proposed, such as applying a sharp confining potential \cite{Menaf2021Qubits,Pawlowski2018ValleyQubit}, creating a lateral heterostructure~\cite{Pawlowski_hetero_2024}, or by the coupling with defect states \cite{Joe2024defects,Krishnan2023defects}. It has been shown that spin–valley locked quantized conductance \cite{Wang2018TMDCQDs,Justin_2023, Pawlowski_Hawrylak_2024} and electrical control over charged excitons \cite{pisoni2018gate} can be achieved experimentally. Furthermore, theoretical studies of electrically defined \ac{TMDC}-based \acp{QD} have shown potential to realize qubits, including spin qubits, valley qubits and hybrid spin–valley qubits \cite{Bjork2004QD,Kormanyos2014QD,Pawlowski2018ValleyQubit,Pawlowski_2019,Pawlowski2021Qubit}. We note that the previous works focused on the $K$-valley states \cite{Pawlowski2018ValleyQubit, Pawlowski_2019, szulakowska2020valley, Menaf2021Qubits, Pawlowski2021Qubit, Miravet2023WSe2QD, Pawlowski_Hawrylak_2024, Miravet_Hawrylak_2024}. In this work, we discuss the possibility of controlling different valley minima using external electric field.

One of the promising heterostructures is composed of MoSe$_2$ and WSe$_2$ monolayers~\cite{rivera2016valley, ciarrocchi2019polarization, seyler2019signatures, Tran_Li_2019, Wang_Mak_2019, jauregui2019electrical, Paik_Deng_2019, Unuchek_Kis_2019, Liu_Zhu_2020, li2020dipolar, Bai2020indirectX, Li_Srivastava_2021, Wang_Zhu_2021, Liu_Lui_2021, brotonsgisbert2021moiretrapped, Ma_Shan_2021, Baek_Gerardot_2021, Wang_Xu_2021, Wilson_Xu_2021, Sun_Kis_2022, Barre_Heinz_2022, Yoon_Sun_2022, Troue_Holleitner_2023, Du_Sun_2023, Zhao_Hogele_2023, Hu_Louie_2023, Campbell_Gerardot_2024, Johnson_Liu_2024, Finley2024moire, Wietek_Chernikov_2024, Huang_Hafezi_2024, FowlerGerace_Butov_2024, Joe_Kim_2024, Peterson_Lukin_2024, Soubelet_Finley_2024, Joe_Kim_2024, Hannachi_Jaziri_2024, Steinhoff_Chernikov_2024}. Having similar lattice constants ($3.288$~\AA~and $3.286$~\AA)~\cite{Wilson_Yoffe_1969}, when stacked using the exfoliation technique, they can create long-range order moir\'e potential \cite{seyler2019signatures,choi2021twist,brotonsgisbert2021moiretrapped,li2021interlayer,Volmer2023ValleyPol,Finley2024moire}, however relaxation/annealing-induced atomic reconstruction can make regions that are commensurate~\cite{Rosenberger_Jonker_2020, Baek_Lee_2023}. In contrast, CVD grown samples show no moir\'e pattern~\cite{Rupp_Hogele_2023}. Analyzed experimentally in recent years, MoSe$_2$/WSe$_2$ heterostructures enable the development of novel optoelectronic \cite{rivera2015observation,rivera2016valley,miller2017long,nagler2017giant,gao2017interlayer,wilson2017hetero,calman2018indirect,hsu2018negative,hanbicki2018double,wang2019optical,jauregui2019electrical,wang2019giant,zhang2019highly,Wang2019indirectX,li2020dipolar,calman2020indirect,sigl2020,Bai2020indirectX,joe2021electrically,Finley2023Photons,Li2023quadrupolar,Yu2023quadrupolar, Mahdikhanysarvejahany2022}, 
valleytronic \cite{Wang2015LightCones,Schaibley2016Valleytronics,ciarrocchi2019polarization}, and quantum computing \cite{Yu2017HeteroQC,Cao2019vdWqubit,merkl2019ultrafast,Kiemle2020ChargedQubit,Purz2021qubits,Miao2021qubit,Montblanch2023quantumTech} devices. 

On the theoretical side, several important advancements have already been achieved. An accurate approach using \ac{DFT} established the band gap of type II: the \ac{CB} minimum from MoSe$_2$ layer, the \ac{VB} maximum from WSe$_{2}$ layer. It is also known that the values of spin-orbit gaps are inherited from monolayers. As usual for \ac{2D} materials, in which the electron-electron interactions are strong, it is less clear whether the (optical) gap extrema are in $K$-points or some other valleys contribute to the physics around the Fermi level, usually $\Gamma$ valley at the \ac{VB} and so called $Q/ \Sigma / \Lambda$ points between $K$ and $\Gamma$ in the \ac{CB}. Several works \cite{Wu_MacDonald_2018b, Gillen_Maultzsch_2018, torun2018interlayer, Donck_Peeters_2018, Wozniak_Kunstmann_2020, Hu_Louie_2023, Li_Peeters_2023, Tenorio_Chaves_2023} studied the problem of the optical properties of this system. While the state-of-the-art \textit{ab initio} methods are invaluable, it is much more convenient to use the low-energy methods. For \acp{TMDC} bilayers several works established how to approach the construction of the \ac{TB} and $k \cdot p$ models~\cite{Wu_MacDonald_2017, Wu_MacDonald_2018, RuizTijerina_Falko_2019, Vitale_Lischner_2021}. Focusing on the \ac{TB} models that offer a good trade-off between accuracy and performance, there are several works discussing the \ac{TB} models for homo-bilayers, e.g. bilayer MoS$_2$~\cite{Cappelluti_Guinea_2013, Zahid_Guo_2013, Roldan_Ordejon_2014, Fang_Kaxiras_2015, Shanavas_Satpathy_2015, Venkateswarlu_TramblydeLaissardiere_2020, Zhang_Yuan_2020}. Surprisingly, we are not aware of any model discussing interactions between two different \ac{TMDC} monolayers on the atomistic level in a full orbital basis, taking into account both $d$ and $p$ orbitals. This problem is non-trivial due to the two seemingly contradictory observations: one being that since the interlayer interactions are weak (\ac{vdW}) it should be possible to understand the interlayer coupling using the \ac{NN} $p_z$ orbitals on chalcogenide atoms; second stating that majority of physics of these materials can be understood in terms of the $d$ orbitals localized on metal atoms. In order to get better understanding, we constructed a full interlayer-interaction model that is consistent with the monolayer basis, consisting of both metal and chalcogen orbitals. 

The paper is organized as follows. In Section~\ref{section:FirstPrinciples} we describe the electronic properties of MoSe$_2$/WSe$_2$ heterostructure using \ac{DFT}. We follow by construction of the \ac{TB} model in Section~\ref{section:TB}. In Section~\ref{section:Efield} we discuss the effect of a perpendicular electric field. Using the TB model, in Section~\ref{section:HeteroQD} we describe a lateral gated quantum dot. We demonstrate that the applied vertical electric field enables control over the \ac{QD} energy spectrum by tuning the valley character of the low-energy states, transitioning between $K$ (2-fold degenerate) and $Q$ (6-fold degenerate) configurations. Finally, in Section~\ref{section:Summary} we summarize our main findings.

%%%%%%%%%%%%%%%%%%%%%%%%%%%%%%%%%%%%%%%%%%%%%%%%%%%%%%%%%%%%%%%%%%%%%%%%%%%%%%%%%%%%%%%%%%

%%%%%%%%%%%%%%%%%%%%%%%%%%%%%%%%%%%%%%% DFT studies %%%%%%%%%%%%%%%%%%%%%%%%%%%%%%%%%%%%%%

%%%%%%%%%%%%%%%%% Fig %%%%%%%%%%%%%%%%
\begin{figure}[t]
\includegraphics[width=0.9\linewidth]{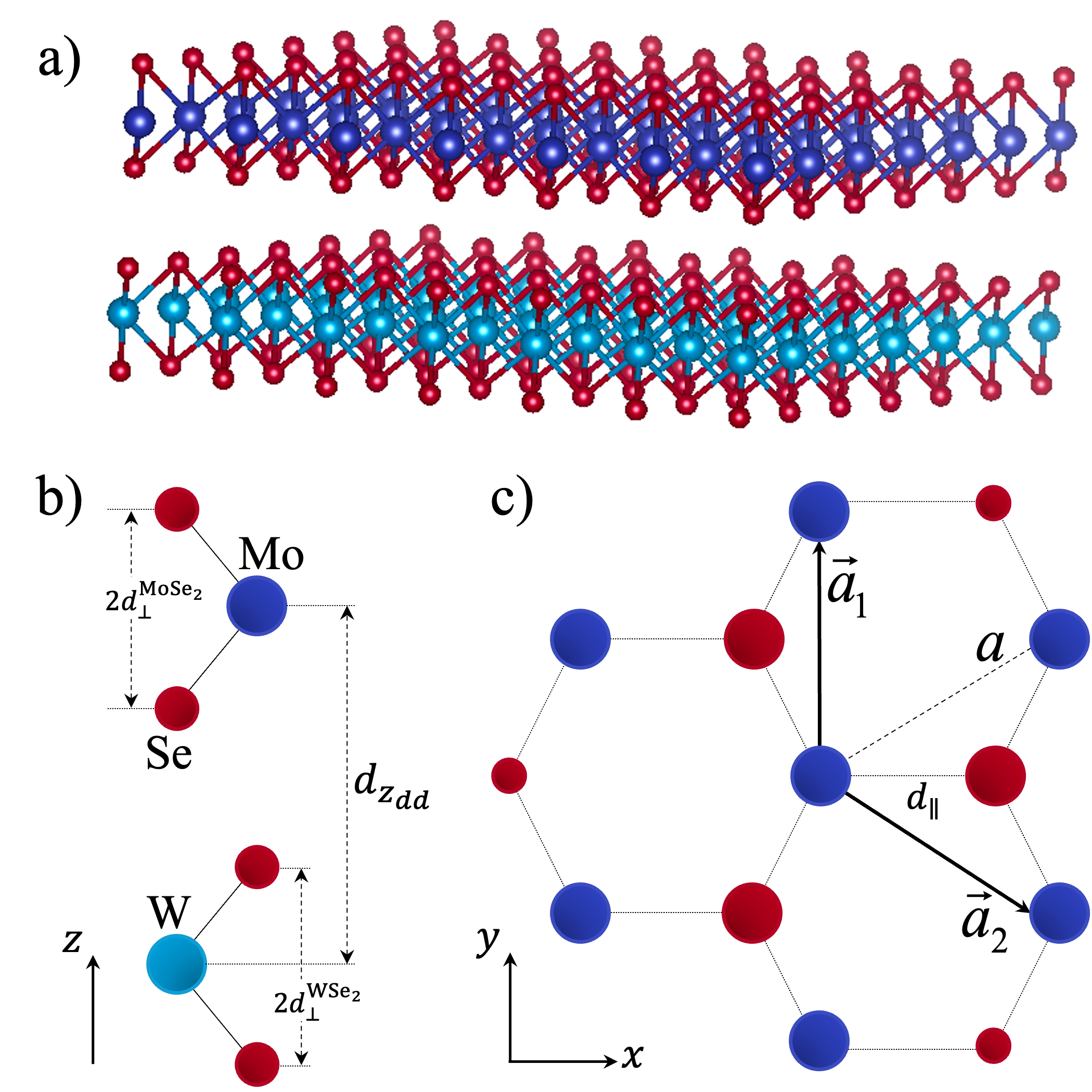}
    \caption{Geometry of the MoSe$_2$/WSe$_2$ heterostructure: (a) \ac{3D}, (b) side, and (c) top views. Top and bottom layers are MoSe$_2$ and WSe$_2$, respectively. Molybdenum (tungsten) metal atoms are represented by dark (light) blue dots, while selenium atoms are shown in red. The interlayer distance has been defined as $d_{z_{dd}}$. The unit cell is shown in (b). Primitive lattice vectors $\vec{a}_{1}$, $\vec{a}_{2}$ are chosen as denoted in (c).}
\label{fig:TMDgeometry}
\end{figure}
%%%%%%%%%%%%%%%%%%%%%%%%%%%%%%%%%%%%%%

%%%%%%%%%%%%%%%%%%%%%%%%%%%%%%%%%%%%%%
\section{\textit{Ab initio} band structure of MoSe$_2$/WSe$_2$ heterostructure}
\label{section:FirstPrinciples}
%%%%%%%%%%%%%%%%%%%%%%%%%%%%%%%%%%%%%%
In this section, we describe the electronic properties of MoSe$_2$/WSe$_2$ from \ac{DFT}. For computational details see Appendix~\ref{section:AppendixDFT}. The \ac{3D} view of the system is shown in Fig.~\ref{fig:TMDgeometry}(a). We study the energetically preferable AB stacking~\cite{Cesare2014stacking}, in which the metal atoms of one layer are located under the chalcogen atoms of the other layer. As presented in Fig.~\ref{fig:TMDgeometry}(b), the \ac{UC} consists of $6$ atoms belonging to the distinct layers. The top view of the system is shown in Fig.~\ref{fig:TMDgeometry}(c), representing the honeycomb lattice. The primitive vectors of the real space lattice are $\vec{a}_{1}=\left(0,a\right)$ and $\vec{a}_2=\left(a\sqrt{3}/2,-a/2\right)$, with $a=d_{||}\sqrt{3}$ being the lattice constant. The $K$ point is then given by $\vec{K}=\left(0,4\pi/3a\right)$. We note that the lattice mismatch in the MoSe$_2$/WSe$_2$ heterostructure is negligibly small, $\Delta a < 0.001$~\AA, thus causing no significant strain. The lattice constant has been set to $a=3.323$~\AA. In order to determine the interlayer distance, we studied the total energy as a function of layers distance \cite{he2014stacking}, obtaining energy minimum for $d_{z_{dd}}=6.400$~\AA. Metal-chalcogen distances $d_{\perp}$ are $2.869$~\AA\  and $2.880$~\AA\  for top and bottom layers, respectively.

%%%%%%%%%%%%%%%%% Fig %%%%%%%%%%%%%%%%
\begin{figure}[t]
\includegraphics[width=\linewidth]{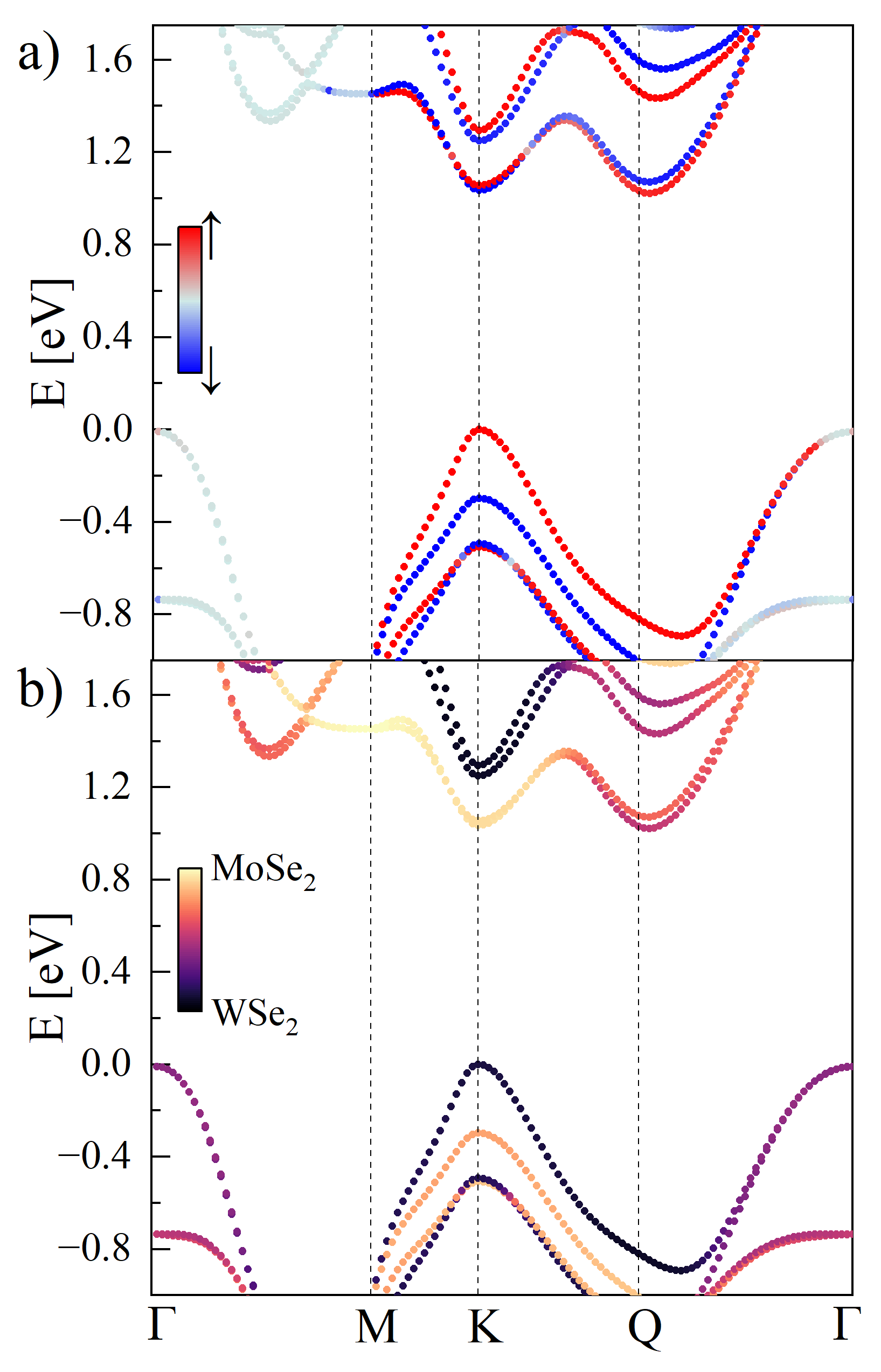}
    \caption{{\textit{Ab initio} electronic structure of the MoSe$_2$/WSe$_2$.} The energy spectrum is presented on the path $\Gamma$-$M$-$K$-$Q$-$\Gamma$. The two panels show (a) spin- and (b) layer-resolved electronic structures. Both spin and layer are decoded by color, where red/blue denotes spin up/down, while yellow/black denotes MoSe$_2$/WSe$_2$ layer, respectively.}
\label{fig:TMDdispersionDFT}
\end{figure}
%%%%%%%%%%%%%%%%%%%%%%%%%%%%%%%%%%%%%%

%%%%%%%%%%%%%%%%%%%%%%%%%%%%%%%%%%%%%%
\subsection{Layer-spin-orbital properties}
%%%%%%%%%%%%%%%%%%%%%%%%%%%%%%%%%%%%%%

Here we analyze the bandstructure of MoSe$_2$/WSe$_2$ heterostructure, as presented in Fig.~\ref{fig:TMDdispersionDFT}. We find that it is characterized by the \ac{CB} minima at $K$ and $Q$ points that are close in energy, separated by $1.8$~meV. The \ac{VB} maxima at $K$ and $\Gamma$ points are separated by 35.5 meV. Fig.~\ref{fig:TMDdispersionDFT}(a) shows the spin-resolved electronic structure. The direct $K$-$K$ energy gap is determined to be $E_g=1.04$~eV, which is smaller compared with separate monoloyers, for which we obtained $E_g^{\text{MoSe}_2}=1.34$ eV and $E_g^{\text{WSe}_2}=1.24$ eV. 
It is worth mentioning that in $K$ points bands around the fundamental gap belong to different layers, as will be discussed later.
The intralayer $K$-valley spin splittings due to the atomic \ac{SOC} in both \ac{CB} and \ac{VB} are $\Delta_{\text{SOC}}^{CB}=12$~meV and $\Delta_{\text{SOC}}^{VB}=211$~meV for MoSe$_2$-localized bands, while for WSe$_2$ $\Delta_{\text{SOC}}^{CB}=30$~meV and $\Delta_{\text{SOC}}^{VB}=473$~meV. The values of spin-orbit splitting are similar to those of single layers. We note that for the direct transition across the fundamental gap $K$-$K$ spin is flipped, while for the transition from $K$ to $Q$ the spin is parallel. 

Furthermore, Fig.~\ref{fig:TMDdispersionDFT}(b) presents the details of the layer contributions. We note the type-II band alignment, where around $K$ valley electrons in the bottom of the \ac{CB} belong to the MoSe$_2$ layer, while holes in the top of the \ac{VB} belong to the WSe$_2$ layer, in agreement with previous works \cite{Wozniak_Kunstmann_2020, sadecka2022inter, bieniek2022nanomaterials, FariaJunior_Fabian_2023}. While around the $K$ valley we observe a strong spin-layer localization for both electrons in \ac{CB} and holes in \ac{VB}, in the $Q$ valley electrons are delocalized between distinct layers, as shown in Fig.~\ref{fig:TMDdispersionDFT}(b). Moreover, detailed analysis of the density $\rho^{n}_{\vec{k}}$ (defined in the Appendix~\ref{section:AppendixDFT}) shows that the spin of the electron in $Q$ is mixed ($\approx 20 \%$ admixture), contrary to $K$ valley.

%%%%%%%%%%%%%%%%% Fig %%%%%%%%%%%%%%%%
\begin{figure}[t]
\includegraphics[width=\linewidth]{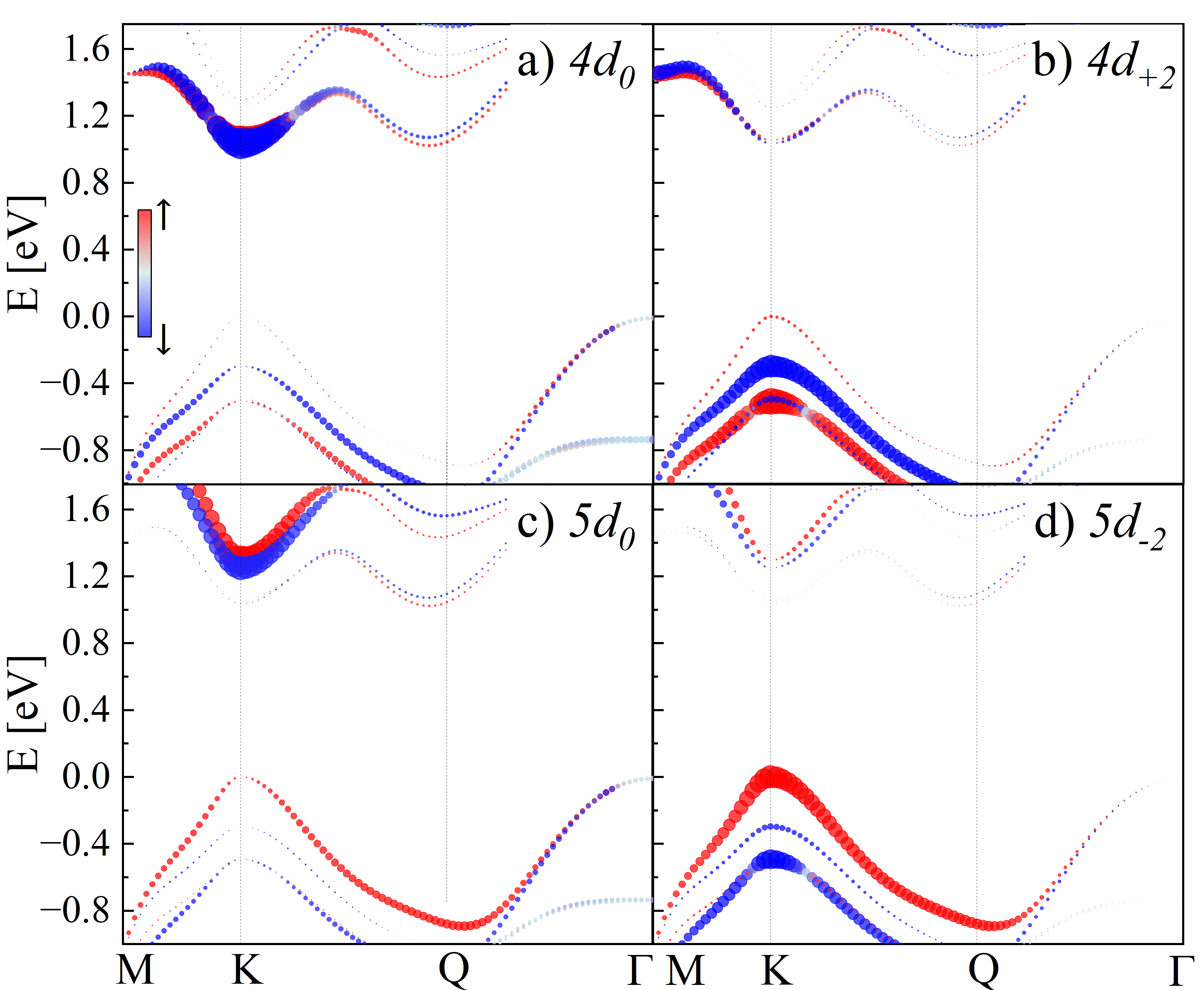}
    \caption{Orbital contribution for the leading orbitals from both layers: (a) $4d_{m=0}$ and (b) $4d_{m=+2}$ from molybdenum, and (c) $5d_{m=0}$ and (d) $5d_{m=-2}$ from tungsten. Color denotes spin.}
\label{fig:OrbitalContribution}
\end{figure}
%%%%%%%%%%%%%%%%%%%%%%%%%%%%%%%%%%%%%%

Using the Kohn-Sham wavefunctions projected onto localized orbitals, the orbital-resolved decomposition of wavefunctions has been performed, see Appendix~\ref{section:AppendixDFT} and Fig.~\ref{fig:OrbitalContribution}. Similar to the case of a single layer of \ac{TMDC}s~\cite{bieniek2022nanomaterials,bieniek2020band}, the main contribution to the bands around the Fermi level comes from the symmetric (even) orbitals. The \ac{CB} is mainly composed of the metal orbitals, $4d_{m=0}$ for Mo and $5d_{m=0}$ for W, depicted in Fig.~\ref{fig:OrbitalContribution}(a,c), while the contribution to the \ac{VB} comes mainly from the $4d_{m=\pm 2}$ orbitals for Mo and $5d_{m=\pm 2}$ for W, shown in Fig.~\ref{fig:OrbitalContribution}(b,d). Due to the AB stacking the dominant orbital contribution around valley $K$ in the valence band is $4d_{m=+2}$ for MoSe$_2$ layer and $5d_{m=-2}$ for WSe$_2$ layer, in line with simple picture of opposite valleys on top of each other (however, this does not mean simple valley-layer locking due to the strong layer localization effect). The orbital contribution around the $Q$ valley is more complex, reflecting the effect of electron delocalization between distinct layers. Detailed values of the orbital contributions to energy bands at high symmetry points are given in Tab.~\ref{tab:OrbContributionDFT} in Appendix~\ref{section:AppendixDFT}.

%%%%%%%%%%%%%%%%%%%%%%%%%%%%%%%%%%%%%%%%%%%%%%%%%%%%%%%%%%%%%%%%%%%%%%%%%%%%%%%%%%%%%%%%%%

%%%%%%%%%%%%%%%%%%%%%%%%%%%%%%%%% Tight-Binding Model %%%%%%%%%%%%%%%%%%%%%%%%%%%%%%%%%%%%

%%%%%%%%%%%%%%%%%%%%%%%%%%%%%%%%%%%%%%
\section{Tight-binding model}
\label{section:TB}
%%%%%%%%%%%%%%%%%%%%%%%%%%%%%%%%%%%%%%

In the following section we construct the \textit{ab initio}-based \ac{TB} model for a type-II \ac{TMDC} heterostructure. This approach allows us to understand the interlayer interactions and the effect of an applied electric field. It also opens the possibility to study heterobilayer-based nanostructures.

%%%%%%%%%%%%%%%%%%%%%%%%%%%%%%%%%%%%%%
\subsection{Orbital basis}
\label{section:TBbasis}
%%%%%%%%%%%%%%%%%%%%%%%%%%%%%%%%%%%%%%

Building on previous works on monolayer \ac{TMDC}s \cite{bieniek2018zeeman, bieniek2020band, Menaf2021Qubits, bieniek2022nanomaterials}, we construct a two-layer, spinfull, even-odd orbital model of the AB stacked heterostructure in \ac{NNN} hopping approximation, for both inter- and intralayer hoppings. The electron wavefunction is defined as a linear combination of localized atomic orbitals $\psi_{l}$ (for details see Appendix~\ref{section:AppendixTBrot}) and has the following form \cite{bieniek2020band}:
%%%%%%%%%%%%%% Eq \ac{TB} WF %%%%%%%%%%%%%%
\begin{align}\label{eq:TBwavefunction}
    \varphi^{p}_{\vec{k}}\left(\vec{r}\right)&= %e^{i\vec{k}\cdot\vec{r}} u^{p}_{\vec{k}}\left(\vec{r}\right) =  
	\frac{1}{\sqrt{N_{U}}} \sum_{i=1}^{N_{U}}\sum_{l=1}^{N_l} e^{i\vec{k}\cdot\vec{U}_{i,l}} A^{p}_{\vec{k},l}\psi_{l}\left(\vec{r}-\vec{U}_{i,l}\right)=\nonumber\\
  &=\sum_{l=1}^{N_l} A^{p}_{\vec{k},l}\phi_{\vec{k},l}(\vec{r}),
\end{align}
%%%%%%%%%%%%%%%%%%%%%%%%%%%%%%%%%%%%%%
where $p$ denotes bands, $\vec{k}$ is a wavevector on the \ac{2D} \ac{BZ}, and $\phi_{\vec{k},l}(\vec{r})$ are the orbital Bloch functions. The number of \acp{UC} is $N_{U}$, while $N_{l}$ corresponds to the number of orbitals on different sublattices. The orbitals are localized at $\vec{U}_{i,l} = \vec{U}_{i}+\vec{\tau}_{l}$,  where $U_i$ are the \ac{UC} coordinates, and $\tau_l$ denotes the $l$-th atomic orbital positions inside of each \ac{UC}. The coefficients $A^{p}_{\vec{k},l}$ are solution to the \ac{TB} Hamiltonian, Eq.~(\ref{eq:HeteroHamiltonian}), $H_\mathrm{hetero}(\vec{k})\vec{A}^{p}_{\vec{k}}=\varepsilon^{p}_{\vec{k}}\vec{A}^{p}_{\vec{k}}$. Based on the \ac{DFT} studies we have taken the $d$ metal atom orbitals with $m=\{0,\pm1,\pm2\}$ and the $p$ chalcogen dimers constructed out of orbitals with $m=\{0,\pm1\}$. Thus, the number of orbitals on metal atom is $5$ and on chalcogen dimer sites is $6$ (including even and odd combinations). 

It is convenient to separate the sum over orbitals $\sum_l$ between two subspaces, even $\mathcal{H}_\mathrm{even}$ and odd $\mathcal{H}_\mathrm{odd}$ with respect to the single layer metal planes. The constructed basis is ordered in the following way: $\{m_d=-2,0, 2, m_p=-1, 0, 1\}$ for even subspace, and $\{m_d=-1, 1, m_p=-1, 0, 1\}$ for the odd subspace. The chalcogen dimers $X_2$ for $p$ orbitals are defined for both subspaces as 
$m_{p_{+1}} = (\ket{m_{p_{+1}}^{X^{(1)}}}\pm\ket{m_{p_{+1}}^{X^{(2)}}})/\sqrt{2}$,
$m_{p_0} = (\ket{m_{p_0}^{X^{(1)}}}\mp\ket{m_{p_0}^{X^{(2)}}})/\sqrt{2}$, and
$m_{p_{-1}} = (\ket{m_{p_{-1}}^{X^{(1)}}}\pm\ket{m_{p_{-1}}^{X^{(2)}}})/\sqrt{2}$,
where the $+$/$-$ sign corresponds to the even/odd subspace and $1$/$2$ index to the top/bottom atom within the $X_2$ dimer. The full spinless even-odd wavefunction of the monolayer can be thus written as $\varphi^{p}_{\vec{k}}\left(\vec{r}\right) = \left[\varphi^{p,\text{ev}}_{\vec{k}}\!\left(\vec{r}\right),\varphi^{p,\text{odd}}_{\vec{k}}\!\left(\vec{r}\right)\right]^T.$ We note that in this chiral basis we need to use linear combinations of standard Slater-Koster \cite{Slater_Koster_1954} hopping integrals.

%%%%%%%%%%%%%%%%%%%%%%%%%%%%%%%%%%%%%%
\subsection{Spinfull monolayer Hamiltonian}
\label{section:monoH}
%%%%%%%%%%%%%%%%%%%%%%%%%%%%%%%%%%%%%%

Monolayer Hamiltonians for both MoSe$_2$ and WSe$_2$ with \ac{SOC} can be written, following previous works, as \cite{bieniek2020band,bieniek2022nanomaterials,Menaf2021Qubits}:
%%%%%%%%%%%%%% Eq H mono %%%%%%%%%%%%
\begin{equation}
    H_{\text{MX$_2$}} =
    \begin{bmatrix}
        H^{\text{ev}}_{\uparrow} & 0 & 0 & H^{\text{ev-odd}}_{\uparrow\downarrow} \\
         & H^{\text{odd}}_{\uparrow} & H^{\text{odd-ev}}_{\uparrow\downarrow} & 0 \\
         & & H^\text{ev}_{\downarrow} & 0 \\
         & & & H^\text{odd}_{\downarrow}
    \end{bmatrix}.
\label{eq:MonoHamiltonian}
\end{equation}
%%%%%%%%%%%%%%%%%%%%%%%%%%%%%%%%%%%%%%
The Hilbert space of $\hat{H}$ defined in Eq.~(\ref{eq:MonoHamiltonian}) is given by $\mathcal{H}_\mathrm{\text{MX$_2$}}=(\mathcal{H}_\mathrm{even}\oplus\mathcal{H}_\mathrm{odd})\otimes\mathcal{H}_\mathrm{spin}$, thus it is characterized by the dimension $(6+5)\times2$, and hence it is represented by a $22\times22$ matrix. The atomic \ac{SOC} has been also included. It is defined as $\hat{H}_\mathrm{SOC}=\sum_{a}\lambda_a/\hbar\,\hat{\mathbf{L}}_a\cdot\hat{\mathbf{S}}_a$~\cite{Menaf2021Qubits}, with $\lambda_a$ depending on the specific atom, $\hat{\mathbf{L}}_a$ being the atomic orbital angular momentum operator, and $\hat{\mathbf{S}}_a$ the spin operator. Further details on constructing the monolayer Hamiltonian can be found in the Appendix~\ref{section:AppendixTBrot}.

%%%%%%%%%%%%%%%%%%%%%%%%%%%%%%%%%%%%%%
\subsection{Heterostructure Hamiltonian}
\label{section:heteroH}
%%%%%%%%%%%%%%%%%%%%%%%%%%%%%%%%%%%%%%

The \ac{TB} Hamiltonian for MoSe$_2$/WSe$_2$ can be written in a block form, emphasizing distinct layers and the interlayer interaction, as:
%%%%%%%%%%%%% Eq Hetero H %%%%%%%%%%%%%
\begin{equation}
    \hat{H}_\mathrm{hetero} = 
    \begin{bmatrix}
			\hat{R}^{\pi}_z H_{\text{MoSe$_2$}} & H_{\text{inter}}\otimes \mathbf{1}_\sigma \\
                                            & H_{\text{WSe$_2$}}
    \end{bmatrix},
\label{eq:HeteroHamiltonian}
\end{equation}
%%%%%%%%%%%%%%%%%%%%%%%%%%%%%%%%%%%%%%
To incorporate AB stacking we choose to rotate the MoSe$_2$ monolayer by $\pi$, thus rotating the \text{MoSe$_2$} Hamiltonian block: $\hat{R}^{\pi}_z H_{\text{MoSe$_2$}}$, in contrast to the normal $H_{\text{WSe$_2$}}$ orientation. Full form of both Hamiltonians is derived in Appendix~\ref{section:AppendixTBrot}, where for clarity we use $\hat{R}^{\pi}_z H_{\text{MoSe$_2$}}=H^{(\pi)}$. The coupling between the distinct layers is denoted by $H_{\text{inter}}$ which is the same for both spin blocks. All in all, the Hilbert space is a sum of $\mathcal{H}_\mathrm{MoSe_2}\oplus\mathcal{H}_\mathrm{WSe_2}$, thus, it is characterized by the total dimension $22+22$.

%%%%%%%%%%%%%%%%%%%%%%%%%%%%%%%%%%%%%%
\subsection{Interlayer coupling}
\label{section:TBfull}
%%%%%%%%%%%%%%%%%%%%%%%%%%%%%%%%%%%%%%

%%%%%%%%%%%%%%%%% Fig %%%%%%%%%%%%%%%%
\begin{figure}[t]
\includegraphics[width=0.9\linewidth]{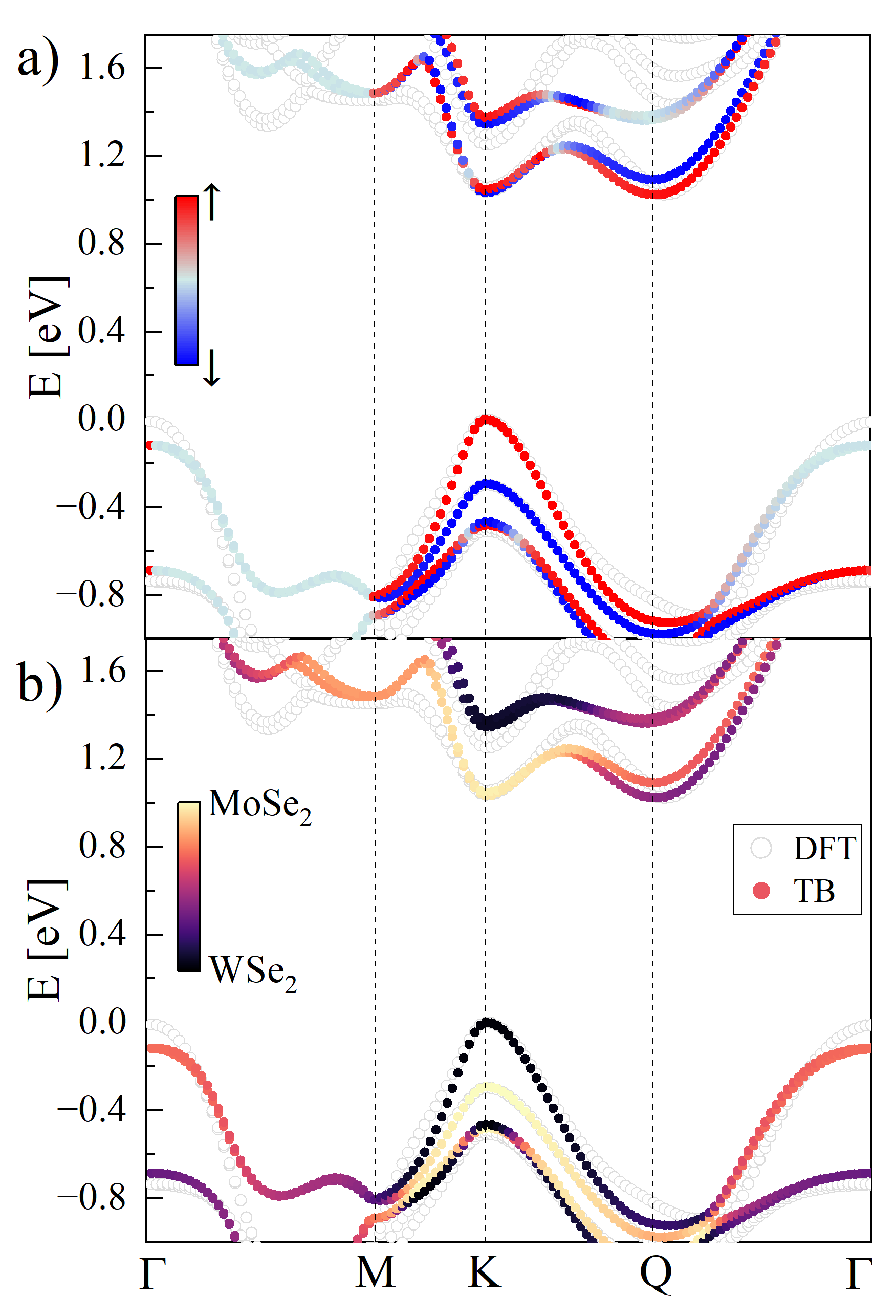}
    \caption{{Electronic structure of the MoSe$_2$/WSe$_2$ in the \ac{TB} model.} Panels present (a) spin and (b) layer resolved band structures. Both spin and layer are represented by color, where red/blue denotes spin up/down, while yellow/black denotes MoSe$_2$/WSe$_2$ layer, respectively. \ac{DFT} results are given by open circles for reference. The parameter set used here is given in Appendix~\ref{section:AppendixTBfull} in Tab.~\ref{tab:SKParametersFull}.}
\label{fig:TMDdispersionTB}
\end{figure}
%%%%%%%%%%%%%%%%%%%%%%%%%%%%%%%%%%%%%%

We now move to construct a \ac{TB} theory of the interlayer interactions. We include up to \acp{NNN} interaction, which, e.g., includes interlayer metal-metal interaction. The coupling between MoSe$_2$ and WSe$_2$ layers is defined in a block form as:
%%%%%%%%%%%%% Eq H inter %%%%%%%%%%%%%
\begin{equation}
    H_{\text{inter}} =
    \begin{bmatrix}
        H_{dd}^\text{ev-ev} & H_{dp}^\text{ev-ev} & H_{dd}^\text{ev-odd} & H_{dp}^\text{ev-odd} \\
        H_{pd}^\text{ev-ev} & H_{pp}^\text{ev-ev} & H_{pd}^\text{ev-odd} & H_{pp}^\text{ev-odd}\\
				H_{dd}^\text{odd-ev} & H_{dp}^\text{odd-ev} & H_{dd}^\text{odd-odd} & H_{dp}^\text{odd-odd} \\
        H_{pd}^\text{odd-ev} & H_{pp}^\text{odd-ev} & H_{pd}^\text{odd-odd} & H_{pp}^\text{odd-odd}
    \end{bmatrix},
\label{eq:InterHamiltonianFull}
\end{equation}
%%%%%%%%%%%%%%%%%%%%%%%%%%%%%%%%%%%%%%
with the interlayer couplings between Mo and W metal atoms ($dd$), chalcogen dimers ($pp$) and metal-chalcogen dimer interactions ($pd$/$dp$). The details of the Hamiltonian~(\ref{eq:InterHamiltonianFull}) can be found in Appendix~\ref{section:AppendixTBfull}.

%%%%%%%%%%%%%%%%%%%%%%%%%%%%%%%%%%%%%%
\subsection{Tight-binding model dispersion and spin-layer-orbital properties}
\label{section:TBresults}
%%%%%%%%%%%%%%%%%%%%%%%%%%%%%%%%%%%%%%

Next we parametrize $26$ intralayer and $9$ interlayer Slater-Koster parameters, as well as $4$ SOC strengths, as listed in Tab.~\ref{tab:SKParametersFull}. This parameter set has been determined by fitting the \ac{TB} model to the \ac{DFT} results, described in Section~\ref{section:FirstPrinciples}, using the \ac{DE} method \cite{Price2005DiffEvol}. We choose the parameters focusing on obtaining the degenerate \ac{CB} minima around $K$ and $Q$ valleys and \ac{VB} maxima around $K$ and $\Gamma$. Moreover, we ensure the correct spin ordering of bands by including an additional term in the \ac{DE} loss function:
\begin{align}
    \mathcal{L}_\mathrm{DE}=&\sum_{p \in \mathcal{P}}\sum_{\vec{k}\in\mathcal{K}} 
    \biggl(
    \lambda_1\left|\varepsilon^{p}_{\vec{k}}-\varepsilon^{p,\mathrm{DFT}}_{\vec{k}}\right| + \nonumber\\ &+\lambda_2\left|\braket{\varphi^{p}_{\vec{k}}|\hat{\sigma}_z|\varphi^{p}_{\vec{k}}}-\sigma^\mathrm{DFT}_z\right|
    \biggr),
\end{align}
where the summation goes over selected subbands $\mathcal{P}$ close to the bandgap, and subset $\mathcal{K}$ of $\vec{k}$-points located nearby the high symmetry points ($K$/$K'$, $Q$, and $\Gamma$) within the \ac{BZ}. The $\lambda_1=1.0$ (eV$^{-1}$) term focuses on \ac{TB} energies $\varepsilon^{p}_{\vec{k}}$  (bands) fitting to the \ac{DFT} results, while the $\lambda_2=0.2$ regularizer tries to enforce the appropriate spin ordering in the \ac{TB} eigenstates $\varphi^{p}_{\vec{k}}$, corresponding to these energies.
 
We obtain the single-particle spectrum and wavefunctions in the \ac{TB} approximation by diagonalizing the Hamiltonian given by Eq.~(\ref{eq:HeteroHamiltonian}). Fig.~\ref{fig:TMDdispersionTB} presents the spin-resolved \ac{TB} electronic structure. The microscopic details of spin and layer contributions stay in agreement with the \textit{ab initio} results. While the \ac{CB} is characterized by a degenerate minima for $K$ and $Q$ valleys, in the \ac{VB} maxima for $K$ and $\Gamma$ are split more than in \ac{DFT}. The direct $K$-$K$ energy gap is of the order of $1.03$~eV. The spin-layer ordering of bands around the energy gap in $K$ point has been captured correctly. Furthermore, while the direct $K$-$K$ transition from the top \ac{VB} to the bottom \ac{CB} is dark due to spin, the indirect transition $K$-$Q$ is bright due to spin, as shown in Fig.~\ref{fig:TMDdispersionTB}(a). The splittings of bands due to the \ac{SOC} are consistent with those from \ac{DFT}. Fig.~\ref{fig:TMDdispersionTB}(b) presents the microscopic details of the layer contribution. We confirm the type-II band alignment in our \ac{TB} model. We note that the effect of the electron delocalization between layers in \ac{CB} for the $Q$ point has been captured correctly.

In the next step, we perform the orbital-resolved decomposition of wavefunctions, checking the leading orbital for each $k$ point and each band. In agreement with the \ac{DFT} results (see Section~\ref{section:FirstPrinciples}), the main orbital contribution to the low-energy bands for the high symmetry points $K$ and $Q$ comes from the symmetric (even) orbitals. While the contribution to the \ac{VB} comes mainly from the metal atom orbitals $4d_{m=0}$ for Mo and $5d_{m=0}$ for W, the \ac{CB} is mainly composed of $4d_{m=\pm 2}$ orbitals for Mo and $5d_{m=\pm 2}$ for W. Further details of orbital contributions to the energy bands at the $K$ and $Q$ points are presented in Appendix~\ref{section:AppendixTBfull} in Tab.~\ref{tab:OrbContributionTB}.

We realize that our full $44\times44$ \ac{TB} model is complicated and might be challenging for implementation. While we leave a systematic reduction of it to a future work, we offer a simplified interlayer coupling analysis, that couples even blocks of orbitals in monolayers. This allows to study even-only \ac{TB} model, which has $6$ orbitals per layer per spin. We also simplified the interlayer interaction to capture most important orbital couplings. Details and behavior of such simplified \ac{TB} model are studied in Appendix~\ref{section:AppendixTBsimple}.

%%%%%%%%%%%%%%%%%%%%%%%%%%%%%%%%%%%%%%
\section{Electric Field Effect}
\label{section:Efield}
%%%%%%%%%%%%%%%%%%%%%%%%%%%%%%%%%%%%%%

In the following section we show that the energetic ordering of valleys can be controlled by applied vertical electric field. We introduce this field to the \ac{TB} model by adding $H_E$ term. The field $E_z$ creates negative and positive voltages $V_E(z)=E_z z$ at the lower and upper \ac{TMDC} layers, e.g., generated by vertical gates placed above and below (substrate) the heterostructure. We set $V_E\left(z=d_{z_{dd}}/2\right)=0$, such that on the metal atoms belonging to the upper layer the voltage is given as $+\frac{1}{2}E_z d_{z_{dd}}/\varepsilon$, while for the metal atoms of the bottom layer the voltage is $-\frac{1}{2}E_z d_{z_{dd}}/\varepsilon$.
We assume typical dielectric screening for TMDCs materials as $\varepsilon\simeq7$~\cite{epsilon_tmdc}.

%%%%%%%%%%%%%%%%%%%%%%%%%%%%%%%%%%%%%%
\subsection{Tight-binding model in electric field}
%%%%%%%%%%%%%%%%%%%%%%%%%%%%%%%%%%%%%%

Following Ref. \cite{Menaf2021Qubits}, we write the electric field block $H_E$ for both layers as (same for both spin components):
%%%%%%%%%%%%%% Eq H mono %%%%%%%%%%%%%
\begin{align}\label{eq:EfieldInclusionGeneral}
    &\hat{H}_E=\frac{1}{\varepsilon}\!
    \begin{bmatrix}
        H_{E}^{\mathrm{MoSe}_2}\otimes\mathbf{1}_\sigma & 0 \\
         0 & H_{E}^{\mathrm{WSe}_2}\otimes\mathbf{1}_\sigma
    \end{bmatrix},\\
    \mathrm{with}\nonumber\\
    &H_{E}^{\mathrm{MoSe}_2}=
    \begin{bmatrix}
        \hat{V}_E^\text{ev} & \hat{V}_E^{\text{ev-odd}} \\
         & \hat{V}_E^\text{odd}
    \end{bmatrix},\;
    H_{E}^{\mathrm{WSe}_2}
    \begin{bmatrix}
         -\hat{V}_E^\text{ev} & \hat{V}_E^{\text{ev-odd}} \\
         & -\hat{V}_E^\text{odd}
    \end{bmatrix}.\nonumber
\end{align}
%%%%%%%%%%%%%%%%%%%%%%%%%%%%%%%%%%%%%%
Here $\hat{V}_E^\mathrm{ev}$ and $\hat{V}_E^\mathrm{odd}$ are just diagonal blocks with the same element on the diagonal $E_z \frac{1}{2}d_{z_{dd}}$, i.e., $\hat{V}_E^\mathrm{ev}=E_z \frac{1}{2}d_{z_{dd}}\,\mathrm{diag}(1,1,1,1,1,1)$,  while for
$\hat{V}_E^\mathrm{odd}=E_z \frac{1}{2}d_{z_{dd}}\,\mathrm{diag}(1,1,1,1,1)$.
The block $\hat{V}_E^{\text{ev-odd}}$ couples even and odd orbital subspace, with $V_{X_2}=\frac{1}{2}(V_{X^{(1)}}-V_{X^{(2)}})$ being the potential difference between chalcogen atoms in the dimer:
%%%%%%%%%%%%%% Eq H mono %%%%%%%%%%%
\begin{equation}
    \hat{V}_E^{\text{ev-odd}} =
    \begin{bmatrix}
        0 & 0 & 0 & 0 & 0 \\
        0 & 0 & 0 & 0 & 0 \\
        0 & 0 & 0 & 0 & 0 \\
        0 & 0 & V_{X_2} & 0 & 0 \\
        0 & 0 & 0 & V_{X_2} & 0 \\
        0 & 0 & 0 & 0 & V_{X_2} \\
    \end{bmatrix}.
\label{eq:EfieldInclusion}
\end{equation}
%%%%%%%%%%%%%%%%%%%%%%%%%%%%%%%%%%%%%%
The transformation from the single atom basis to our, i.e., the one using dimers, is described in Appendix~\ref{section:AppendixEfield}.

%%%%%%%%%%%%%%%%%%%%%%%%%%%%%%%%%%%%%%
\subsection{Effect of electric field on heterostructure: \ac{DFT} and \ac{TB} comparison}
\label{section:Efield_dft_vs_tb}
%%%%%%%%%%%%%%%%%%%%%%%%%%%%%%%%%%%%%%

%%%%%%%%%%%%%%%%% Fig %%%%%%%%%%%%%%%%
\begin{figure*}[t]
\includegraphics[width=\textwidth]{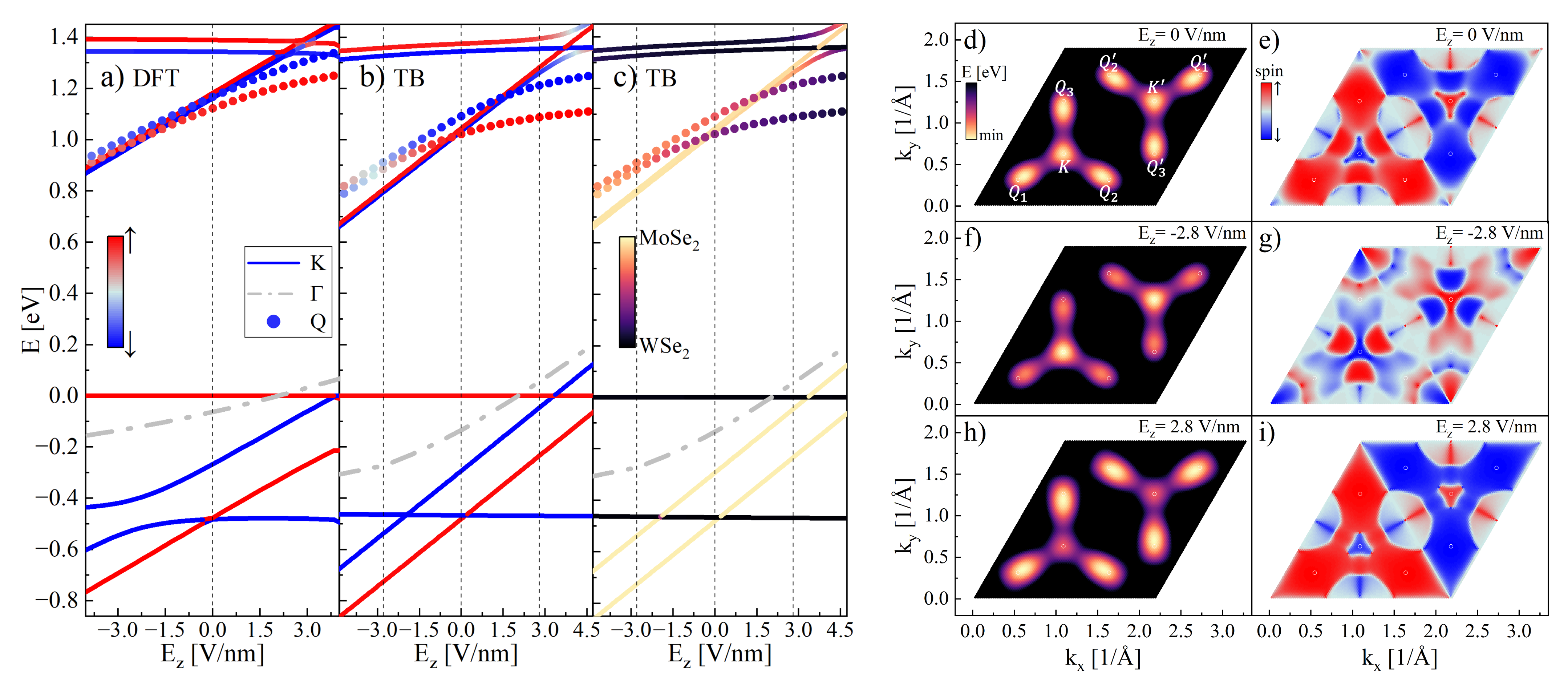}
    \caption{{The effect of a vertical electric field on the electronic structure.} (a-c) Energies of selected bands are presented as a function of the applied $E_z$, following Ref.~\cite{FariaJunior_Fabian_2023}, for $K$, $Q$ and $\Gamma$. Schemes presented for both (a) \ac{DFT} results and (b,c) \ac{TB} ones. Different type of lines represent $K$, $Q$ and $\Gamma$ valleys. Colors correspond to (a,b) spin, and (c) layer compositions, respectively. Due to the degeneracy of \ac{VB} in the $\Gamma$ point, the spin-/layer-resolved notation was not introduced. The vertical dashed lines denote to the electric field $0$ and $\pm2.8$ V/nm. (d,f,h) Energy and (e,g,i) spin color maps (textures) of the lowest \ac{CB} presented on the whole \ac{BZ} for 3 different values of applied electric field: (d,e) $E_z=0$ V/nm, (f,g) $E_z=-2.8$ V/nm, and (h,i) $E_z=2.8$ V/nm, respectively. White circles correspond to the selected high symmetry points, denoted on the plots.}
\label{fig:ElectricFieldTB}
\end{figure*}
%%%%%%%%%%%%%%%%%%%%%%%%%%%%%%%%%%%%%%

Valley $K$, $Q$, and $\Gamma$ states evolution as a function of electric field $E_z$ is presented in Fig.~\ref{fig:ElectricFieldTB}. We obtain satisfying agreement between the \ac{DFT} results -- Fig.~\ref{fig:ElectricFieldTB}(a), and the \ac{TB} model -- Fig.~\ref{fig:ElectricFieldTB}(b,c). To set the stage, we set the top of the \ac{VB} in $K$ point to 0 as a reference independently for each value of the electric field. For both \ac{DFT} and \ac{TB} we observe a clear trends in both \ac{VB} and CB. For large negative values of $E_z$ we observe \ac{VB} maximum at $K$ and \ac{CB} minimum also at $K$. When electric field is increased, \ac{CB} minimum switches between $K$ and $Q$ valleys. Similar effect is observed in VB, where \ac{VB} maximum switches between $K$ and $\Gamma$ valleys. This establishes MoSe$_2$/WSe$_2$ heterostructure as an interesting multi-valley system in which occupation of different type of valleys can be controlled using moderate electric field. 

Focusing now on spin properties, we note that the applied electric field does not change the spin orientation for $K$ valley for both \ac{VB} and \ac{CB}, as shown in Fig.~\ref{fig:ElectricFieldTB}(b). Contrary to that, for negative applied $E_z$ the spin-mixing effect at the $Q$ point is observed. The values of \ac{SOC} splittings in $K$ point of the \ac{CB}s remain almost constant, while in the \ac{VB}s they are strongly affected by the applied field. In the $Q$ point the spin-splitting between the two \ac{CB}s increases with positive $E_z$. Moreover, we notice the flip of spin across the gap, i.e. the lowest \ac{CB} in the $K$ point giving spin-dark, momentum-bright optical transition become spin-bright, momentum dark as the $E_z$ increases.

%Qualitatively the \ac{TB} results capture those of the \ac{DFT}. Nevertheless, there is a feature that the \ac{TB} fails to reproduce: the strong anticrossing in the \ac{VB} at $K$. However, it appears that the simplified model introduced in the Appendix~\ref{section:AppendixTBfull} is able to mimic such mixing -- see Fig.~\ref{fig:ElectricFieldTBsimple}.

In the next step we analyze evolution of the layer contribution, accessible in \ac{TB} -- see Fig.~\ref{fig:ElectricFieldTB}(c). First we note that electric field couples and mixes \ac{CB} much stronger than \ac{VB}. Focusing on \ac{CB}, the $K$ point comes mainly from MoSe$_2$, but for large $E_z$ it can be strongly admixed by WSe$_2$ layer. For the $Q$ valley, the layer contribution is mixed already at $E_z=0$ and can be tuned continuously with the field applied. We note that the occupation of the \ac{CB} $K$ and $Q$ valleys implies not only valley switching, but also switching between layer-localized to layer-delocalized type of bands. 

Now we move to the analysis of \acp{CB} in $E_z$ across the whole \ac{BZ}, summarized in Fig.~\ref{fig:ElectricFieldTB}(d-i). We start with the system with no electric field applied, as presented in Fig.~\ref{fig:ElectricFieldTB}(d,e). In accordance with spin-valley locking, the two non-equivalent valleys $K$ and $K'$ differ by spin, as presented Fig.~\ref{fig:ElectricFieldTB}(e). In the \ac{CB} each of them is surrounded by three $Q$ points, creating a system of 6 non-equivalent $Q$ points in the \ac{BZ}. Also, $K$ and $Q$ valleys are almost degenerate in energy, as can be observed in Fig.~\ref{fig:ElectricFieldTB}(d). Next we apply a moderate electric field $E_z=-2.8$~V/nm -- see Fig.~\ref{fig:ElectricFieldTB}(f,g). We observe breaking of the degeneracy between the energy minima for $K$ and $Q$ valleys. Furthermore, in Fig.~\ref{fig:ElectricFieldTB}(g) we observe the change of spin for the $Q$ valleys, which stays in agreement with the flip of the two bottom \acp{CB} around $Q$ observed in Fig.~\ref{fig:ElectricFieldTB}(b). Finally, for electric field $E_z=2.8$~V/nm -- Fig.~\ref{fig:ElectricFieldTB}(h,i), we observe the effect of breaking the energetic degeneracy in favor of $Q$ valleys. This allows for $E_z$-field controlled occupation of the $Q$-valleys.

It should be also noted that a similar \ac{CB} landscape composed of three valleys close to each $K$-point emerges in a bilayer graphene subjected to a perpendicular electric field, as a result of interplay between trigonal warping and $E_z$-field induced  band gap~\cite{Varlet2014}. These triple degenerate states have been studied theoretically~\cite{Knothe2020,AlbertHawrylak2024BLGTW} and experimentally~\cite{Overweg2018,Garreis2021,Seiler2024}.

%%%%%%%%%%%%%%%%% Fig %%%%%%%%%%%%%%%%
\begin{figure}[tb]
\includegraphics[width=\linewidth]{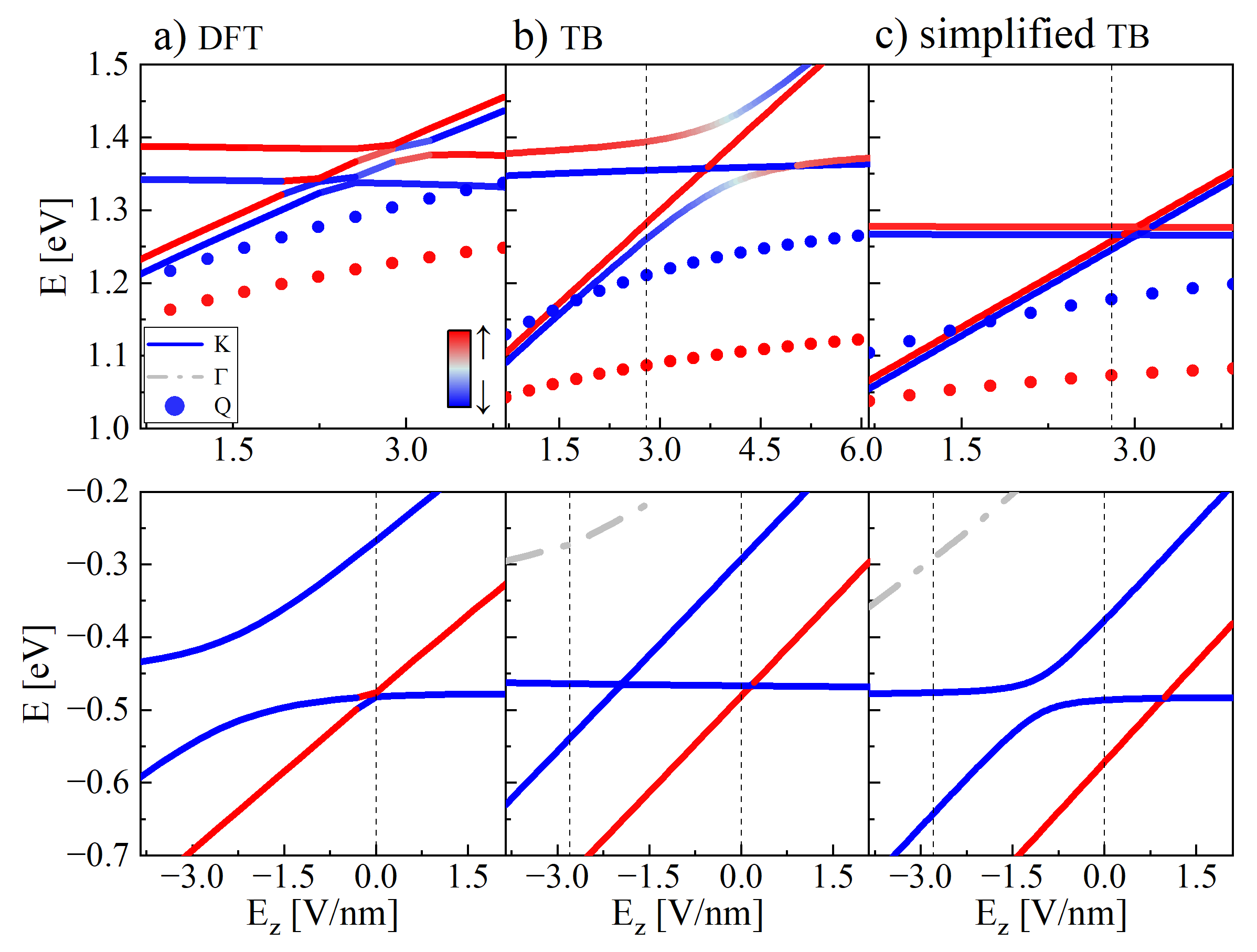}
    \caption{Analysis of anti-crossing of $K$-point states in electric field. Top panel shows zoom into conduction band and lower one for valence band. Column (a) shows band evolution obtained using \ac{DFT}, (b) - full \ac{TB} model and (c) - simplified \ac{TB} model. Note that color denotes spin and electric field on $x$-axis is different between top and bottom panels.}
\label{fig:anticross}
\end{figure}
%%%%%%%%%%%%%%%%%%%%%%%%%%%%%%%%%%%%%%

%%%%%%%%%%%%%%%%% Fig %%%%%%%%%%%%%%%%
\begin{figure*}[tb]
\includegraphics[width=\textwidth]{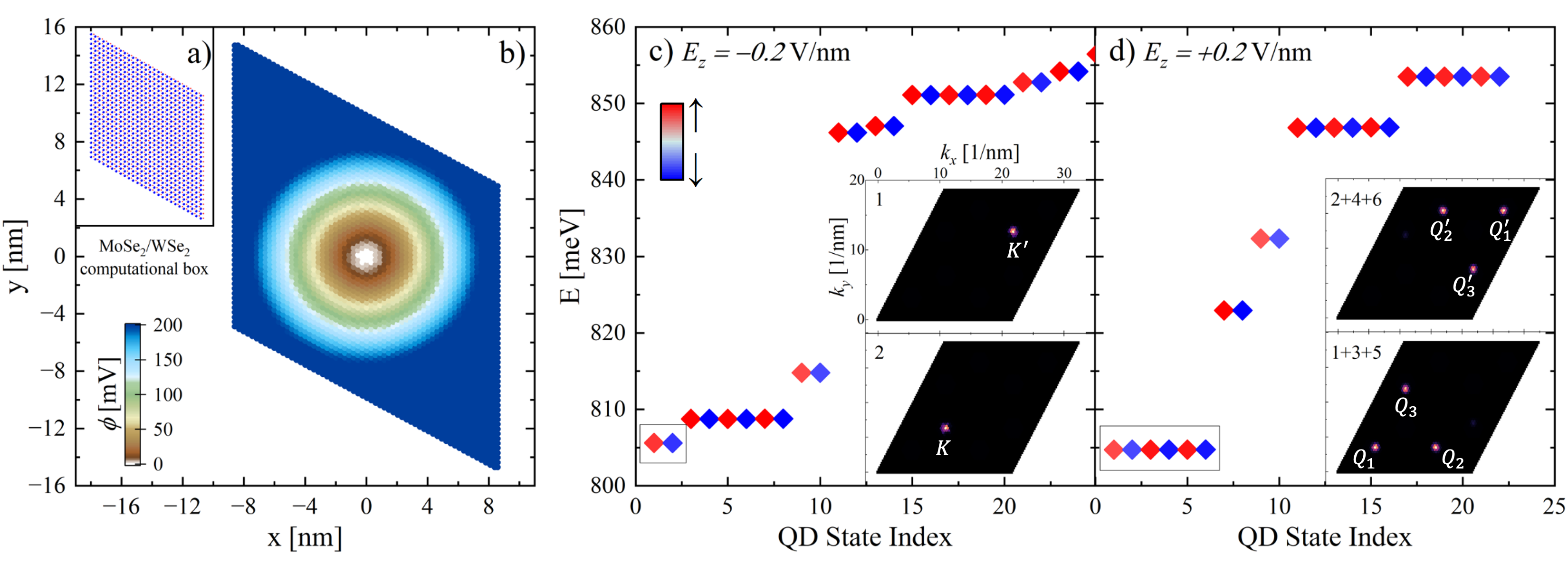}
    \caption{(a) Scheme of the computational box in real space. (b) Electric potential $\phi$ profile of a Gaussian \ac{QD}. (c) Spin-resolved \ac{QD} energy spectrum for vertical electric field $E_z=-0.2$~V/nm and \ac{CB} minimum located in the $K$ valleys. (d) Analogous spectrum for $E_z=+0.2$~V/nm, for which \ac{CB} minima are in $Q$ points. Spin is indicated by color. The insets in panel (c) show density for the first $|\psi_1|^{2}$ and the second $|\psi_2|^{2}$ state, and the insets in panel (d) present two sums over states in the lowest shell in a specific spin subspace ($|\psi_1|^{2}+|\psi_3|^{2}+|\psi_5|^{2}$ and $|\psi_2|^{2}+|\psi_4|^{2}+|\psi_6|^{2}$, respectively).}
\label{fig:QDEfield}
\end{figure*}
%%%%%%%%%%%%%%%%%%%%%%%%%%%%%%%%%%%%%%

%As mentioned earlier, the $K$ point energy evolution as a function of the electric field for states below (above) valence (conduction) band shows discrepancies between the full \ac{TB} model and \ac{DFT}, as presented in the bottom panels of Fig.~\ref{fig:anticross}(a,b). For the \ac{VB}-1 (one band below the \ac{VB} maximum) and the \ac{VB}-2 for $E_z \approx -1.5$~V/nm \ac{DFT} predicts strong anti-crossing for the same spins (blue). This effect is not captured in the full \ac{TB} and we attribute it to the different layer and orbital contributions obtained with the Slater-Koster parametrization used. Since the orbital contribution is modified in the simplified \ac{TB} model (introduced in the Appendix~\ref{section:AppendixTBsimple}) due to the different parametrization (see Tab.~\ref{tab:SKParametersFull}, the anticrossing of bands can be observed, as shown in Fig.~\ref{fig:anticross}(c).
As previously noted, the energy evolution of the $K$ point as a function of the electric field for states below (above) valence (conduction) band exhibits discrepancies between the full \ac{TB} model and \ac{DFT}, as shown in the bottom panels of Fig.~\ref{fig:anticross}(a,b). For the \ac{VB}-1 (one band below the \ac{VB} maximum) and \ac{VB}-2 states at $E_z \approx -1.5$~V/nm \ac{DFT} predicts a strong anticrossing for same-spin states (blue), an effect not captured by the full \ac{TB} model. We attribute this discrepancy to differences in the layer and orbital contributions arising from the Slater-Koster parametrization used. However, in the simplified \ac{TB} model (introduced in Appendix~\ref{section:AppendixTBsimple}), where the parametrization is different (see Tab.~\ref{tab:SKParametersFull}), the anticrossing of bands is reproduced, as illustrated in Fig.~\ref{fig:anticross}(c).
%Interestingly, this particular anti-crossing can be observed for simplified \ac{TB} model (introduced in the Appendix~\ref{section:AppendixTBsimple}), shown in Fig.~\ref{fig:anticross}(c), in which, presumably, smaller amount of fitting parameters allowed for easier optimization of layer/orbital contribution to valence bands. In contrast, an anti-crossing in conduction band at $K$ between \ac{CB} and \ac{CB}+3 is observed for $E_z\approx 1.8$~(V/nm) for full \ac{TB} model, in accordance with \ac{DFT} result -- see top panels of Fig.~\ref{fig:anticross}. However, this effect is missing in the simplified \ac{TB}. 

%%%%%%%%%%%%%%%%%%%%%%%%%%%%%%%%%%%%%%
\section{Electrically tunable laterally gated quantum dot}
\label{section:HeteroQD}
%%%%%%%%%%%%%%%%%%%%%%%%%%%%%%%%%%%%%%
We have demonstrated how electric field perpendicular to the MoSe$_2$/WSe$_2$ layers shifts the relative energies of different valleys. In this section, we leverage this effect in a lateral gate-defined \ac{QD}. We investigate this system using the approach outlined in Refs.~\cite{Bieniek_Hawrylak_2020, Menaf2021Qubits, Miravet2023WSe2QD, Saleem2023Exciton, Korkusinski2023BLGQD, Sadecka2024Trion, Pawlowski_Hawrylak_2024}. We begin with the MoSe$_2$/WSe$_2$ computational rhombus, which serves as the basis for defining Bloch functions in $k$-space. These functions are subsequently employed to describe the states of the gated \ac{QD} and to analyze the effect of the applied vertical electric field, presented in Fig.~\ref{fig:QDEfield}.

To describe the \ac{QD} we start by creating the rhomboidal real-space computational box consisting of MoSe$_2$ and WSe$_2$ atoms, depicted in Fig.~\ref{fig:QDEfield}(a). We impose the periodic boundary conditions connecting the opposite edges of the rhombus, thus giving us a set of allowed $\vec{k}$-vectors. We expand the electron wavefunction in the basis of Bloch functions $\varphi^{p}_{\vec{k}}( \vec{r})$, see Eq.~(\ref{eq:TBwavefunction}), built of WSe$_2$ and MoSe$_2$ atomic orbitals. For each $\vec{k}$ wavevector we diagonalize the Hamiltonian given in Eq.~(\ref{eq:HeteroHamiltonian}) and obtain bulk electronic bands $\varepsilon^{p}_{\vec{k}}$ and eigenvectors $A^{p}_{\vec{k},l}$. We confine electrons in a gate-defined lateral Gaussian potential $V_\mathrm{QD}=U_{QD}(1-\exp(-r^2/\sigma_{QD}^2/2))$, presented in Fig.~\ref{fig:QDEfield}(b). The basis of the band states of the bulk system are functions
%%%%%%%%%%%%%%%%%%%%%%%%%%%%%%%%%%%%%%%%%%%%%%%%%%%%%%%%%%
$\varphi^{p}_{\vec{k}}( \vec{r})=\sum_l A^{p}_{\vec{k},l}\phi_{\vec{k},l}(\vec{r})$,
%%%%%%%%%%%%%%%%%%%%%%%%%%%%%%%%%%%%%%%%%%%%%%%%%%%%%%%%%%
where $l$ denotes the orbital, $p$ corresponds to the band index, and $\phi_{\vec{k},l}(\vec{r})$ are the orbital Bloch functions build from the localized Slater orbitals. The \ac{QD} states, expanded in the basis of low-energy band states $\varphi^{p}_{\vec{k}}$, are defined as: 
%%%%%%%%%%%%%%%%%%%%%%%% QD States %%%%%%%%%%%%%%%%%%%%%%%
\begin{equation}
    \Phi^{s}\left(\vec{r}\right) = \sum_p\sum_{\vec{k}} B^{s,p}_{\vec{k}} \varphi^{p}_{\vec{k}}(\vec{r}),
\label{eq:QDwavefunction}
\end{equation}
%%%%%%%%%%%%%%%%%%%%%%%%%%%%%%%%%%%%%%%%%%%%%%%%%%%%%%%%%%
where the summation is carried over the bands $p$ and the wave vectors $\vec{k}$ defined by the computational rhombus. The Schr\"odinger equation gives the integral equation for the amplitudes $B^{s,p}_{\vec{k}}$:
%%%%%%%%%%%%%%%%%%%%% eq for QD States %%%%%%%%%%%%%%%%%%%
\begin{equation}
    \varepsilon^{p}_{\vec{k}} B^{s,p}_{\vec{k}} + \sum_{p',\vec{k'}} \braket{\varphi^{p}_{\vec{k}}|V_\mathrm{QD}|\varphi^{p'}_{\vec{k'}}} B^{s,p'}_{\vec{k'}} = \epsilon^{s} B^{s,p}_{\vec{k}}\;,
    \label{eq:QDequation}
\end{equation}
%%%%%%%%%%%%%%%%%%%%%%%%%%%%%%%%%%%%%%%%%%%%%%%%%%%%%%%%%%
with $\varepsilon^{p}_{\vec{k}}$ denoting the single-particle energies of the heterostructure Hamiltonian described by Eq.~(\ref{eq:HeteroHamiltonian}). The coupling between the band states due to the confinement potential $V_\mathrm{QD}$ is determined by the matrix elements given as:
%%%%%%%%%%%%%% eq for QD potential V^eff %%%%%%%%%%%%%%%%%
\begin{equation}
    \braket{\varphi^{p}_{\vec{k}}|V_\mathrm{QD}|\varphi^{p'}_{\vec{k}'}} =\sum_l \left(A^{p}_{\vec{k},l}\right)^{\ast}\!A^{p'}_{\vec{k'},l}  e^{i\left(\vec{k}'-\vec{k}\right)\cdot \vec{\tau}_l} V_{\vec{k},\vec{k}',l}\;.
    \label{eq:QDpotential}
\end{equation}
%%%%%%%%%%%%%%%%%%%%%%%%%%%%%%%%%%%%%%%%%%%%%%%%%%%%%%%%%%
Here $\vec{\tau}_l$ corresponds to the position of orbitals within a \ac{UC}, while $V_{\vec{k},\vec{k}',l}$ is the Fourier transform of the \ac{QD} potential. 

The eigenenergies and eigenstates of the \ac{QD} system, denoted as $\epsilon^s$ and $\Phi^s$, respectively, are determined by solving Eq.~(\ref{eq:QDequation}). The \ac{QD} potential is modeled as a Gaussian potential characterized by a width of $\sigma_\mathrm{QD}=5$~nm and an amplitude of $U_\mathrm{QD} = 300$~mV. In our calculations, we restrict $p, p'$ to the two lowest \acp{CB}. Figure~\ref{fig:QDEfield}(c,d) illustrates the spin-resolved \ac{QD} energy spectrum for two directions of an applied vertical electric field $E_z$. When a negative electric field is applied, electrons localize in the $K$ and $K'$ valleys, resulting in a doubly degenerate shell of lowest-energy states, as shown in Fig.~\ref{fig:QDEfield}(c). The wavefunctions for these states are confined within the $K$/$K'$ valleys, as illustrated in the insets of Fig.~\ref{fig:QDEfield}(c). The second shell originates from the secondary (higher) \ac{CB} minimum at the 6~$Q$ points. The third shell, also doubly degenerate, arises from the spin-split bands at the $K$ points. The degeneracies of higher-energy shells can be understood in terms of the Fock-Darwin spectrum, including doubly degenerate $2p$-$2p$ states at the $K$ points (four states in total) and $2p$-$2p$ states from the $Q$ valleys ($2\times 6$ states in total), and further states following the same pattern.

Applying a positive electric field shifts electron localization from the $K$ to the $Q$ valleys, resulting in a 6-fold degenerate low-energy shell of \ac{QD} states, as shown in Fig.~\ref{fig:QDEfield}(d). In this configuration, the wavefunctions for each spin subspace form a superposition across all three $Q$ valleys, as illustrated in the insets of Fig.~\ref{fig:QDEfield}(d). The second and third shells are doubly degenerate, originating from spin-split bands around the $K$ points. This behavior aligns with the reversal of valley character ($Q$ vs. $K$) in successive shells, which governs their degeneracy.

%%%%%%%%%%%%%%%%%%%%%%%%%%%%%%%%%%%%%%%%%%%%%%%%%%%%%%%%%%%%%%%%%%%%%%%%%%%%%%%%%%%%%%%%%%

%%%%%%%%%%%%%%%%%%%%%%%%%%%%%%%%%%%%%%%%% Summary %%%%%%%%%%%%%%%%%%%%%%%%%%%%%%%%%%%%%%%%

\section{Summary}
\label{section:Summary}
In summary, we developed here a theory of a laterally gated quantum dot in an electrically tunable WSe$_2$/MoSe$_2$ heterostructure. We employed \textit{ab initio} methods and derived a tight binding model to determine the tunability of valley contributions to the electronic structure of the WSe$_2$/MoSe$_2$ heterostructure with electric field. Our analysis revealed a type-II band alignment with energetically close \ac{CB} minima at the $K$ and $Q$ points. We characterized the electronic properties, including the microscopic details of the layer, spin, and orbital contributions. Building on these results, we developed and parametrized a tight-binding model to describe the heterostructure, capturing the interplay between the two types of valleys in the \ac{CB}. Furthermore, we explored the effect of a vertical electric field on the low-energy valley character of the \ac{CB} minimum and \ac{VB} maximum, validating our findings against the \ac{DFT} calculations. Using the \ac{TB} model, we studied a laterally gated quantum dot. We demonstrated that the \ac{QD} energy spectrum can be effectively tuned by an applied vertical electric field, enabling control over the valley character of the low-energy states between $K$-valley 2-fold and $Q$-valley 6-fold degenerate configurations.
%%%%%%%%%%%%%%%%%%%%%%%%%%%%%%%%%%%%%%%%%%%%%%%%%%%%%%%%%%%%%%%%%%%%%%%%%%%%%%%%%%%%%%%%%%

%%%%%%%%%%%%%%%%%%%%%%%%%%%%%%%%%%%%% Acknowledgments %%%%%%%%%%%%%%%%%%%%%%%%%%%%%%%%%%%%

\section{Acknowledgments}
We thank  M. Korkusiński, J. Fabian, D. Miravet and Y. Saleem for fruitful discussions. This research was supported by the National Science Centre, Poland, under Grant No. 2021/43/D/ST3/01989, and Digital Research Alliance Canada with computing resources,
as well as by PLGrid (WCSS) high-performance computing infrastructure. P.H. acknowledges financial support from the Quantum Sensors Challenge Program QSP078 and Applied Quantum Computing AQC004 Challenge Programs at the National Research Council of Canada, NSERC Discovery Grant No. RGPIN- 2019-05714, and University of Ottawa Research Chair in Quantum Theory of Materials, Nanostructures, and Devices. P.E.F.J. acknowledges the financial support of the Deutsche Forschungsgemeinschaft (DFG, German Research Foundation) SFB 1277 (Project ID No. 314695032, Projects No. B07 and No. B11) and SPP 2244 (Project ID No. 443416183).

%%%%%%%%%%%%%%%%%%%%%%%%%%%%%%%%%%%%%%%%%%%%%%%%%%%%%%%%%%%%%%%%%%%%%%%%%%%%%%%%%%%%%%%%%%

%%%%%%%%%%%%%%%%%%%%%%%%%%%%%%%%%%%%%%%% Appendix A %%%%%%%%%%%%%%%%%%%%%%%%%%%%%%%%%%%%%%
\section*{Appendix}
\appendix
\section{DFT study details}
\label{section:AppendixDFT}
%%%%%%%%%%%%%%%%%%%%%%%%%%%%%%%%%%%%%%

The energies and wavefunctions have been calculated with the use of \ac{DFT} methods,  as implemented in Abinit \cite{Gonze2002}. \ac{PBE} parametrization of the \ac{GGA} for both the exchange-correlation potentials \cite{Perdew1996} and the \ac{PAW} method \cite{Blochl1994} has been used. Both the atomistic spin-orbit interaction and the \ac{vdW} interctions have been taken into account, where for \ac{vdW} interactions the \ac{DFT}-D3 exchange-correlation functional \cite{Grimme2010} has been applied. The \ac{PBE} parametrization has been used with a plane-wave cutoff set to 160~Ha, the energy cutoff of 80~Ha, and the $k$-points grid $16\times 16\times 1$, respectively. Vacuum between the primitive cells along $z$-direction has been set to 40~\AA. We have determined the lattice constant to be $a=3.323$~\AA, while the chalcogen atoms distances in $z$ direction are $d_{XX}^{\text{MoSe}_2}=3.398$~\AA~and $d_{XX}^{\text{WSe}_2}=3.360$~\AA, and metal-chalcogen distances are $d_{MX}^{\text{MoSe}_2}=2.542$~\AA~and $d_{MX}^{\text{WSe}_2}=2.550$~\AA, respectively. We note that throughout this paper the eigenenergies are shifted such that the Fermi energy is always $E_F=0$. Electric field results were obtained as outlined in Ref. \cite{FariaJunior_Fabian_2023}.

Subsequently, we study the general properties of the \ac{DFT} Kohn–Sham wavefunctions $\varphi^{n,\mathrm{DFT}}_{\vec{k}}$, where $n$ denotes bands and $k$ denotes points from the reciprocal space. We define a density $\rho^{n}_{\vec{k}}(z) = 
\iint_{UC} | \varphi^{n,DFT}_{\vec{k}} \left( x,y,z \right) |^2 dx dy$ allowing us to determine the microscopic details of the leading spin and layer contributions, described in Section~\ref{section:FirstPrinciples}. %We have determined the spin-splittings around $K$ valley and presented them in Tab.~\ref{tab:SOCandEMA}.

In the next step, we calculate the orbital contributions to the energy bands of the MoSe$_2$/WSe$_2$ heterostructure using the Kohn-Sham wavefunctions projected onto Slater orbitals. The Slater-like localized orbitals are
%%%%%%%%%% Eq Slater Orbital %%%%%%%%%
\begin{equation}
   \psi_{l}\left(\vec{r}\right)= \psi_{nLm}\left(\vec{r}\right) = R_{n,L}\left(r\right) Y_{Lm}\left(\theta,\phi\right),
\label{eq:SlaterOrbital}
\end{equation}
where $n$, $L$ and $m$ denote the main, orbital angular momentum and magnetic quantum numbers. The radial function $R_{n,L}$ has been approximated as
\begin{equation}
    R_{n,L}\left(r\right) = \frac{\left(2\zeta_{n,L}\right)^{n+\frac{1}{2}}}{\sqrt{\left(2n\right)!}} r^{n-1} e^{-\zeta_{n,L} r},
\label{eq:RadialFunction}
\end{equation}
where the Slater parameters $\zeta_{n,L}$ being taken for the isolated atom model, specifically (in inverse Bohr radius units $[a_{0}^{-1}]$) $\zeta_{4p,Se}=2.0718$ , $\zeta_{4d,Mo}=3.111$ and $\zeta_{5d,W}=3.3484$ \cite{Clementi_Raimondi_1963, Clementi_Reinhardt_1967}. Spherical harmonics $Y_{L,\mu}$ are given by the formula
\begin{equation}
    Y_{L,m}\left(\theta,\phi\right) = \sqrt{\frac{2L+1}{4\pi}\frac{\left(L-m\right)!}{\left(L+m\right)!}} P^m_L \left(\cos\theta\right) e^{im\phi}.
\label{eq:SphericalHarmonics}
\end{equation}
In the above expression $P^m_L$ are the associated Legendre polynomials with Condon-Shortley phase $\left(-1\right)^m$ included.
%%%%%%%%%%%%%%%%%%%%%%%%%%%%%%%%%%%%%%
For molybdenum and tungsten atoms we have $L=2$ and $m\in\{-2,0,+2\}$. For selenium atoms with $L=1$ we have $m\in\{-1,0,+1\}$. We note that for each layer top and bottom Se atoms have been treated as a dimer Se$_2$. We calculate the overlap of Konh-Sham wavefuntions with \ac{TB}-localized orbitals in spheres surrounding atoms, which radius is half of the distance to the nearest atom. All orbital compositions are normalized to $100$\% within valence shell orbitals space. All the main orbital contributions to the top \ac{VB} and bottom \ac{CB} are presented in Tab.~\ref{tab:OrbContributionDFT}.

%%%%%%%%%%%%%%% Tab %%%%%%%%%%%%%%%%%%
\begin{table}[t]
\centering 
\caption{Orbital contribution from \ac{DFT} for the even subspace presented for the atomic orbitals with the main contribution to the top \ac{VB} and bottom \ac{CB}. Numbers are represented in \% for a given high symmetry point. Sign '-' represents 0\%/0\% contribution.}
\vspace{2mm}
\setlength\cellspacetoplimit{2pt}
\setlength\cellspacebottomlimit{4pt}
\begin{tabular}{ | O{P{1.2cm}} O{P{1.2cm}} | O{P{1.35cm}} O{P{1.35cm}} O{P{1.35cm}} O{P{1.35cm}} | } \hline\hline

                             &           & K, \ac{VB}   & K, \ac{CB} & Q, \ac{CB} & $\Gamma$, \ac{VB}  \\ 
     & & $\downarrow$/$\uparrow$ & $\downarrow$/$\uparrow$ & $\downarrow$/$\uparrow$ & $\downarrow$/$\uparrow$ \\ \hline\hline
                                 & $4p_{-1}$ & -        & 0\%/20\% & 3\%/1\%  & -         \\
    Se$_2^{(l_{\text{top}})}$    & $4p_{0}$  & -        & -        & 4\%/1\%  & 6\%/6\%   \\ 
                                 & $4p_{+1}$ & 4\%/0\%  & -        & 6\%/2\%  & -         \\\hline
                                 & $4d_{-2}$ & -        & - & 14\%/5\% & 1\%/0\%   \\
    Mo                           & $4d_{0}$  & -        & 0\%/73\% & 8\%/2\%  & 11\%/11\%  \\
                                 & $4d_{+2}$ & 8\%/0\%  & -        & 3\%/1\%  & 0\%/1\%   \\\hline\hline
                                 & $4p_{-1}$ & 27\%/0\% & -        & 6\%/1\%  & -         \\
    Se$_2^{(l_{\text{bottom}})}$ & $4p_{0}$  & -        & -        & 3\%/1\%  & 6\%/6\%   \\
                                 & $4p_{+1}$ & -        & -        & 2\%/1\%  & -         \\\hline
                                 & $5d_{-2}$ & 53\%/0\% & -        & 2\%/2\%  & 1\%/0\%         \\
    W                            & $5d_{0}$  & -        & 2\%/0\%  & 5\%/1\%  & 11\%/11\% \\
                                 & $5d_{+2}$ & -        & 0\%/1\%  & 11\%/1\% & 0\%/1\%   \\\hline\hline

\end{tabular}
\label{tab:OrbContributionDFT}
\end{table}
%%%%%%%%%%%%%%%%%%%%%%%%%%%%%%%%%%%%%%

%%%%%%%%%%%%%%%%%%%%%%%%%%%%%%%%%%%%%%%% Appendix B %%%%%%%%%%%%%%%%%%%%%%%%%%%%%%%%%%%%%%
\section{Details on the single layer tight-binding Hamiltonian}
\label{section:AppendixTBrot}
%%%%%%%%%%%%%%%%%%%%%%%%%%%%%%%%%%%%%%

In Section~\ref{section:TBfull} we have presented the heterostructure Hamiltonian given by Eq.~(\ref{eq:HeteroHamiltonian}), that contains the blocks corresponding to the MoSe$_{2}$ and WSe$_{2}$ monolayers and the interlayer coupling between them, $H_{\text{inter}}$. Both monolayer \ac{TB} models correspond to the minimal $p^3d^5$ \textit{ab initio}-based \ac{TB} described in our previous works \cite{bieniek2018zeeman,bieniek2020band,bieniek2022nanomaterials,Menaf2021Qubits}. However, we have highlighted the need to modify one of those blocks due to the AB-stacking presented in Fig.~\ref{fig:TMDgeometry}(b) and considered throughout this study. This geometry corresponds to the 180$^\circ$ in-plane rotation of one layer with respect to the other. Hence, the \ac{TB} Hamiltonian for one of the \ac{TMDC} monolayers has to be modified accordingly, which results in the difference between $H_{\text{MoSe}_2}^{\text{rot}}$ and $H_{\text{WSe}_2}$ defined in Eq.~(\ref{eq:HeteroHamiltonian}). 

In the case of non-rotated layer (denoted here by the superscript $(0)$), the monolayer \ac{NNN} Hamiltonian can be defined in a block form for both even and odd subspace as \cite{bieniek2018zeeman,bieniek2020band,bieniek2022nanomaterials}:
%%%%%%%%%% Eq H mono 0deg %%%%%%%%%%%%
\begin{equation}
    H^{(0)}=
    \begin{bmatrix}
        H_{M-M}^{(0),\text{ev}} & H_{M-X_2}^{(0),\text{ev}} & 0 & 0 \\
         & H_{X_{2}-X_{2}}^{(0),\text{ev}} & 0 & 0 \\
				  & & H_{M-M}^{(0),\text{odd}} & H_{M-X_2}^{(0),\text{odd}} \\
					& & & H_{X_2-X_2}^{(0),\text{odd}} 
    \end{bmatrix}.
\label{eq:fullTBmono}
\end{equation}
Furthermore, the matrix describing metal-metal \ac{NNN} interactions in the even subspace is given by:
\begin{equation}
    H_{M-M}^{(0),\text{ev}}=
    \begin{bmatrix}
    ^{E_{m_{_{d}}=-2}}_{{+}W_{1}g_{0}(\vec{k})}& W_{3}g_{2}(\vec{k}) & W_{4}g_{4}(\vec{k}) \\
    & ^{E_{m_{_{d}}=0}}_{{+}W_{2}g_{0}(\vec{k})} & W_{3}g_{2}(\vec{k})\\ 
    &  & ^{E_{m_{_{d}}=2}}_{{+}W_{1}g_{0}(\vec{k})} 
    \end{bmatrix},
\label{eq:TBmonoMMeven}
\end{equation}
and in the odd subspace by:
\begin{equation}
    H_{M-M}^{(0),\text{odd}}=
    \begin{bmatrix}
    ^{E_{m_{_{d}}=-1}}_{{+}W_{8}g_{0}(\vec{k})} & -W_{9}g_{2}(\vec{k}) \\
    & ^{E_{m_{_{d}}=+1}}_{{+}W_{8}g_{0}(\vec{k})} 
    \end{bmatrix}.
\label{eq:TBmonoMModd}
\end{equation}
Next, the corresponding matrix describing $X_{2}$-$X_{2}$ dimer interactions in the even subspace is defined as:
\begin{equation}
    H_{X_2-X_2}^{(0),\text{ev}}=
    \begin{bmatrix}
    ^{E_{m_{_{p}}=-1}}_{{+}W_{5}g_{0}(\vec{k})} & 0 & -W_{7}g_{2}(\vec{k}) \\
    & ^{E_{m_{_{p}}=0}}_{{+}W_{6}g_{0}(\vec{k})} & 0 \\
    & & ^{E_{m_{_{p}}=1}}_{{+}W_{5}g_{0}(\vec{k})}
   \end{bmatrix},
\label{eq:TBmonoXXeven}
\end{equation}
end in the odd subspace:
\begin{equation}
    H_{X_2-X_2}^{(0),\text{odd}}=
    \begin{bmatrix}
    ^{E^{\text{odd}}_{m_{_{p}}=-1}}_{{+}W_{5}g_{0}(\vec{k})} & 0 & -W_{7}g_{2}(\vec{k}) \\
    & ^{E^{\text{odd}}_{m_{_{p}}=0}}_{{+}W_{6}g_{0}(\vec{k})} & 0 \\
    & & ^{E^{\text{odd}}_{m_{_{p}}=1}}_{{+}W_{5}g_{0}(\vec{k})}
   \end{bmatrix},
\label{eq:TBmonoXXodd}
\end{equation}
Finally, the metal-dimer tunneling is described in the even subspace as:
\begin{equation}
    H_{M-X_2}^{(0),\text{ev}}=
    \begin{bmatrix}
    V_{1}f_{-1}(\vec{k}) & -V_{2}f_{0}(\vec{k}) & V_{3}f_{1}(\vec{k}) \\
    -V_{4}f_{0}(\vec{k}) & -V_{5}f_{1}(\vec{k}) & V_{4}f_{-1}(\vec{k}) \\
    -V_{3}f_{1}(\vec{k}) & -V_{2}f_{-1}(\vec{k}) & -V_{1}f_{0}(\vec{k}) 
   \end{bmatrix},
\label{eq:TBmonoMXeven}
\end{equation}
and in the odd subspace:
\begin{equation}
    H_{M-X_2}^{(0),\text{odd}}=
    \begin{bmatrix}
    -V_{6}f_{+1}(\vec{k}) & -V_{8}f_{-1}(\vec{k}) &  V_{7}f_{0}(\vec{k}) \\
     V_{7}f_{-1}(\vec{k}) &  V_{8}f_{0}(\vec{k})  & -V_{6}f_{+1}(\vec{k})
   \end{bmatrix}.
\label{eq:TBmonoMXodd}
\end{equation}
%%%%%%%%%%%%%%%%%%%%%%%%%%%%%%%%%%%%%%
In the monolayer Hamiltonian given by Eq.~(\ref{eq:fullTBmono}) the matrix elements of \ac{NN} tunneling are expressed by $k$-independent parameters $V_{i}$:
%%%%%%%%%%%%%%%% Eq V_i %%%%%%%%%%%%%%
{\allowdisplaybreaks
\begin{align}
V_{1}&=
    \frac{1}{\sqrt{2}}\frac{d_{\|}}{d}\left[\frac{\sqrt{3}}{2}\left(\frac{d^{2}_{\bot}}{d^{2}}-1\right)V_{dp\sigma}-\left(\frac{d^{2}_{\bot}}{d^{2}}+1\right)V_{dp\pi} \right], \nonumber\\
V_{2}&=
    \frac{1}{2}\left(\frac{d_{\|}}{d}\right)^{2}\frac{d_{\bot}}{d}\left[\sqrt{3}V_{dp\sigma}-2V_{dp\pi} \right],\nonumber\\
V_{3}&=
    \frac{1}{\sqrt{2}}\left(\frac{d_{\|}}{d}\right)^{3}\left[\frac{\sqrt{3}}{2}V_{dp\sigma}-V_{dp\pi}\right],\nonumber\\
V_{4}&=
    \frac{1}{2}\frac{d_{\|}}{d}\left[\left(3\frac{d^{2}_{\bot}}{d^{2}}-1\right)V_{dp\sigma}-2\sqrt{3}\frac{d^{2}_{\bot}}{d^{2}}V_{dp\pi} \right],\nonumber\\
V_{5}&=
    \frac{1}{\sqrt{2}}\frac{d_{\bot}}{d}\left[\left(3\frac{d^{2}_{\bot}}{d^{2}}-1\right)V_{dp\sigma}-2\sqrt{3}\left(\frac{d^{2}_{\bot}}{d^{2}}-1\right)V_{dp\pi} \right],\nonumber\\
V_{6}&=
    \frac{1}{\sqrt{2}}\frac{d_\bot}{d} \left[\frac{d_\|^2}{d^3} \left(\sqrt{3}V_{dp\sigma}-2V_{dp\pi}\right)+2V_{dp\pi} \right],\nonumber\\
V_{7}&=
    \frac{1}{\sqrt{2}}\frac{d_\bot d_\|^2}{d^3} \left(\sqrt{3}V_{dp\sigma}-2V_{dp\pi}\right),\nonumber\\
V_{8}&=
    \frac{d_\|}{d} \left[\frac{d_\bot^2}{d^2} \left(\sqrt{3}V_{dp\sigma}-2V_{dp\pi}\right)+V_{dp\pi} \right],
\end{align}
}
%%%%%%%%%%%%%%%%%%%%%%%%%%%%%%%%%%%%%%
and $k$-dependent functions $f_{i}$:
%%%%%%%%%%%%%%%% Eq f_i %%%%%%%%%%%%%%
{\allowdisplaybreaks
\begin{align}
    \begin{split}
    f_{-1}(\vec{k})=e^{ik_{x}d_\|}&
    +e^{-ik_{x}d_\|/2}e^{ i\sqrt{3}k_{y}d_\|/2}e^{ i2\pi/3}\\&
    +e^{-ik_{x}d_\|/2}e^{-i\sqrt{3}k_{y}d_\|/2}e^{-i2\pi/3},
    \end{split} \nonumber\\
    \begin{split}
    f_{0}(\vec{k})=e^{ik_{x}d_\|}&
    +e^{-ik_{x}d_\|/2}e^{ i\sqrt{3}k_{y}d_\|/2}e^{-i2\pi/3}\\&
    +e^{-ik_{x}d_\|/2}e^{-i\sqrt{3}k_{y}d_\|/2}e^{ i2\pi/3},
    \end{split} \nonumber\\
    \begin{split}
    f_{+1}(\vec{k})=e^{ik_{x}d_\|}&
    +e^{-ik_{x}d_\|/2}e^{ i\sqrt{3}k_{y}/2}\\&
    +e^{-ik_{x}d_\|/2}e^{-i\sqrt{3}k_{y}/2}.
    \end{split}
\end{align}
}
%%%%%%%%%%%%%%%%%%%%%%%%%%%%%%%%%%%%%%
The parameters of the \ac{NNN} tunneling are given by $k$-independent terms $W_{i}$:
%%%%%%%%%%%%%%%% Eq W_i %%%%%%%%%%%%%%
{\allowdisplaybreaks
\begin{align}
&W_{1}=\frac{1}{8}\left(3V_{dd\sigma}+4V_{dd\pi}+V_{dd\delta}\right),\nonumber\\
&W_{2}=\frac{1}{4}\left(V_{dd\sigma}+3V_{dd\delta}\right),\nonumber\\
&W_{3}=-\frac{\sqrt{3}}{4\sqrt{2}}\left(V_{dd\sigma}-V_{dd\delta}\right),\nonumber\\
&W_{4}=\frac{1}{8}\left(3V_{dd\sigma}-4V_{dd\pi}+V_{dd\delta}\right),\nonumber\\
&W_{5}=\frac{1}{2}\left(V_{pp\sigma}+V_{pp\pi}\right),\nonumber\\
&W_{6}=V_{pp\pi},\nonumber\\ 
&W_{7}=\frac{1}{2}\left(V_{pp\sigma}-V_{pp\pi}\right),\nonumber\\
&W_{8}=\frac{1}{2}\left(V_{dd\pi}+V_{dd\delta}\right),\nonumber\\
&W_{9}=\frac{1}{2}\left(V_{dd\pi}-V_{dd\delta}\right), 
\end{align}
}
%%%%%%%%%%%%%%%%%%%%%%%%%%%%%%%%%%%%%%
and $k$-dependent functions $g_{i}$
%%%%%%%%%%%%%%%% Eq g_i %%%%%%%%%%%%%%
{\allowdisplaybreaks
\begin{align}
    \begin{split}
    g_{0}(\vec{k})= & 4\cos{\left(3k_{x}d_\|/2\right)} \cos{\left(\sqrt{3}k_{y}d_\|/2\right)} +\\
    &2\cos{\left(\sqrt{3}k_{y}d_\|\right)},
    \end{split} \nonumber\\
    \begin{split}
    g_{2}(\vec{k})= & 2\cos{\left(3k_{x}d_\|/2+\sqrt{3}k_{y}d_\|/2 \right)e^{i\pi/3}} +\\
    &2\cos{\left(3k_{x}d_\|/2-\sqrt{3}k_{y}d_\|/2 \right)e^{-i\pi/3}} + \\ 
		&-2\cos{\left(\sqrt{3}k_{y}d_\|\right)},
    \end{split} \nonumber\\
    \begin{split}
    g_{4}(\vec{k})= & 2\cos{\left(3k_{x}d_\|/2+\sqrt{3}k_{y}d_\|/2 \right)e^{i2\pi/3}} +\\
    &2\cos{\left(3k_{x}d_\|/2-\sqrt{3}k_{y}d_\|/2 \right)e^{-i2\pi/3}} + \\
		&2\cos{\left(\sqrt{3}k_{y}d_\|\right)}.
    \end{split}
\end{align}
}
%%%%%%%%%%%%%%%%%%%%%%%%%%%%%%%%%%%%%%

Having fully defined the non-rotated monolayer \ac{TMDC} \ac{NNN} \ac{TB} model, we now move to describing the differences that appear in the Hamiltonian due to the $\pi$ in-plane rotation of one layer with respect to the other. We note that the rotation does not change neither the formulas defining the $\vec{k}$-dependent functions $f$ and $g$, nor the definitions of amplitudes $V$ and $W$. However, it results in the change of the sign in $\vec{k}$ coordinates for the \ac{NN} couplings ($\vec{k}\rightarrow-\vec{k}$ for the metal-chalcogen dimer couplings) and does introduce a $(-1)$ factor in front of the particular \ac{NN} Hamiltonian matrix elements (the metal-chalcogen dimer couplings). Full Hamiltonian for rotated monolayer reads
%%%%%%%%%% Eq H mono 180deg %%%%%%%%%%%%
\begin{equation}
    H^{(\pi)}=
    \begin{bmatrix}
        H_{M-M}^{(0),\text{ev}} & H_{M-X_2}^{(\pi),\text{ev}} & 0 & 0 \\
         & H_{X_{2}-X_{2}}^{(0),\text{ev}} & 0 & 0 \\
				  & & H_{M-M}^{(0),\text{odd}} & H_{M-X_2}^{(\pi),\text{odd}} \\
					& & & H_{X_2-X_2}^{(0),\text{odd}} 
    \end{bmatrix}.
\label{eq:fullTBmonoRot}
\end{equation} 
Below we redefine explicitly the $M$-$X_2$ matrix elements in the rotated \ac{TMDC} monolayer Hamiltonian $H^{(\pi)}$ in the even subspace:
%%%%%%%%%% Eq H mono 180deg %%%%%%%%%%
\begin{equation}
    H_{M-X_2}^{(\pi),\text{ev}}=
    \begin{bmatrix}
    -V_{1}f_{-1}(-\vec{k}) & -V_{2}f_{0}(-\vec{k})  & -V_{3}f_{1}(-\vec{k}) \\
     V_{4}f_{0}(-\vec{k})  & -V_{5}f_{1}(-\vec{k})  & -V_{4}f_{-1}(-\vec{k}) \\
     V_{3}f_{1}(-\vec{k})  & -V_{2}f_{-1}(-\vec{k}) & V_{1}f_{0}(-\vec{k}) 
   \end{bmatrix},
\label{eq:TBmonoMXeven180}
\end{equation}
and in the odd subspace:
\begin{equation}
    H_{M-X_2}^{(\pi),\text{odd}}=
    \begin{bmatrix}
    -V_{6}f_{+1}(-\vec{k}) &  V_{8}f_{-1}(-\vec{k}) &  V_{7}f_{0}(-\vec{k}) \\
     V_{7}f_{-1}(-\vec{k}) & -V_{8}f_{0}(-\vec{k})  & -V_{6}f_{+1}(-\vec{k})
   \end{bmatrix},
\label{eq:TBmonoMXodd180}
\end{equation}
%%%%%%%%%%%%%%%%%%%%%%%%%%%%%%%%%%%%%%
respectively.

Next, we discuss the \ac{SOC} Hamiltonian, given by Eq.~\ref{eq:MonoHamiltonian}. The non-zero matrix elements of $H^{\text{ev}}_{\uparrow}$ block are diagonal and given by $\mathrm{diag}\left(-\lambda_{M}, 0, \lambda_{M}, -1/2\lambda_{X_{2}}, 0, 1/2\lambda_{X_{2}} \right)$. The corresponding odd block $H^{\text{odd}}_{\uparrow}$ is also diagonal $\mathrm{diag}\left(-1/2\lambda_{M}, 1/2\lambda_{M}, -1/2\lambda_{X_{2}},0, 1/2\lambda_{X_{2}} \right)$. Blocks with opposite spin have opposite signs. Even-odd subspace coupling elements are $H^\text{ev-odd}_{\uparrow\downarrow}$ and $H^\text{odd-ev}_{\uparrow\downarrow}$. First block has four non-zero elements given by: 
\begin{itemize}
\item $\bra{m_{d}=-2\uparrow,\text{ev}}\hat{H}_\mathrm{SOC}\ket{m_{d}=-1\downarrow,\text{odd}}=\lambda_{M}$, 
\item 
$\bra{m_{d}=0\uparrow,\text{ev}}\hat{H}_\mathrm{SOC}\ket{m_{d}=1\downarrow,\text{odd}}=\sqrt{3/2}\lambda_{M}$, 
\item 
$\bra{m_{p}=-1\uparrow,\text{ev}}\hat{H}_\mathrm{SOC}\ket{m_{p}=0\downarrow,\text{odd}}=\sqrt{2}/2\lambda_{X_{2}}$,
\item 
$\bra{m_{p}=0\uparrow,\text{ev}}\hat{H}_\mathrm{SOC}\ket{m_{p}=1\downarrow,\text{odd}}=\sqrt{2}/2\lambda_{X_{2}}$.
\end{itemize}
The non-zero elements of the second block $H^\text{odd-ev}_{\uparrow\downarrow}$ are:
\begin{itemize}
\item 
$\bra{m_{d}=-1\uparrow,\text{odd}}\hat{H}_\mathrm{SOC}\ket{m_{d}=0\downarrow,\text{ev}}=\sqrt{3/2}\lambda_{M}$, 
\item
$\bra{m_{d}=1\uparrow,\text{odd}}\hat{H}_\mathrm{SOC}\ket{m_{d}=2\downarrow,\text{ev}}=\lambda_{M}$, 
\item
$\bra{m_{p}=-1\uparrow,\text{odd}}\hat{H}_\mathrm{SOC}\ket{m_{p}=0\downarrow,\text{ev}}=\sqrt{2}/2\lambda_{X_{2}}$,
\item
$\bra{m_{p}=0\uparrow,\text{odd}}\hat{H}_\mathrm{SOC}\ket{m_{p}=1\downarrow,\text{ev}}=\sqrt{2}/2\lambda_{X_{2}}$.
\end{itemize}

%%%%%%%%%%%%%%%%%%%%%%%%%%%%%%%%%%%%%%%% Appendix C %%%%%%%%%%%%%%%%%%%%%%%%%%%%%%%%%%%%%%
\section{Full interlayer tight-binding Hamiltonian}
\label{section:AppendixTBfull}
%%%%%%%%%%%%%%%%%%%%%%%%%%%%%%%%%%%%%%

In the following section we present the details of the \ac{TB} model for interlayer interaction. We define the interlayer Hamiltonian in both even and odd subspace, as well as include the mixing between them, hence the $H_{\text{inter}}$ can be written in the form given by Eq.~(\ref{eq:InterHamiltonianFull}). However, we note the following symmetries between particular couplings:
%%%%%%%%%%%%% Eq H inter %%%%%%%%%%%%%
%{\allowdisplaybreaks
\begin{align}
    &H^{\text{ev-ev}}_{pp}=H^{\text{ev-odd}}_{pp}=
    -H^{\text{odd-odd}}_{pp}=-H^{\text{odd-ev}}_{pp}=H_{pp},\nonumber\\
    &H^{\text{ev-ev}}_{pd}=-H^{\text{odd-odd}}_{pd}=H_{pd}^{(1)},\nonumber\\
    &H^{\text{ev-odd}}_{pd}=-H^{\text{odd-ev}}_{pd}=H_{pd}^{(2)},\nonumber\\
    &H^{\text{ev-ev}}_{dp}=H^{\text{ev-odd}}_{dp}=H_{dp}^{(1)},\nonumber\\
    &H^{\text{odd-odd}}_{dp}=H^{\text{odd-ev}}_{dp}=H_{dp}^{(2)},
\label{eq:InterHamiltonianFullSymm}
\end{align}
%}
%%%%%%%%%%%%%%%%%%%%%%%%%%%%%%%%%%%%%%
thus allowing us to write rewrite Eq.~(\ref{eq:InterHamiltonianFull}) in the following form:
%%%%%%%%%%%%% Eq H inter %%%%%%%%%%%%%
\begin{equation}
H_{\text{inter}} =
\begin{bmatrix}
    H^{(1)}_{dd} & H_{dp}^{(1)}& H^{(2)}_{dd} & H_{dp}^{(1)}\\
    H_{pd}^{(1)} & H_{pp} & H_{pd}^{(2)} & H_{pp} \\
    H^{(3)}_{dd} & H_{dp}^{(2)} & H^{(4)}_{dd} & H_{dp}^{(2)} \\
    -H_{pd}^{(1)} & -H_{pp} & -H_{pd}^{(2)} & -H_{pp}
\end{bmatrix}.
\label{eq:InterHamiltonianFullExpandSimple}
\end{equation}
%%%%%%%%%%%%%%%%%%%%%%%%%%%%%%%%%%%%%%
For the $p$-$p$ couplings, corresponding blocks are defined as:
%%%%%%%%%%%%% Eq Hab inter %%%%%%%%%%%
\begin{equation}
\begin{aligned}
H_{pp} &=
\begin{bmatrix}
    T_{1}f_{+1}(\vec{k}) & -T_{2}f_{-1}(\vec{k}) & T_{3}f_{0}(\vec{k})\\
    T_{2} f_{0}(\vec{k}) & T_{4}f_{+1}(\vec{k}) & -T_{2}f_{-1}(\vec{k}) \\
    T_{3}f_{-1}(\vec{k}) & T_{2}f_{0}(\vec{k}) & T_{1}f_{+1}(\vec{k})
\end{bmatrix}
\end{aligned}
\end{equation}
For the $p$-$d$ and $d$-$p$ couplings we have:
%%%%%%%%%%%%%%%%%%%%%%%%%%%%%%%%%%%%%%
{\allowdisplaybreaks
\begin{align}
H_{pd}^{(1)} &=
\begin{bmatrix}
    0 & 0     & 0\\
    0 & T_{5} & 0 \\
    0 & 0     & 0
\end{bmatrix}, \hspace{6mm}
%%%%%%%%%%%%%%%%%%%%%%%%%%%%%%%%%%%%%%
H_{dp}^{(1)} =
\begin{bmatrix}
    0 & 0     & 0\\
    0 & T_{7} & 0 \\
    0 & 0     & 0
\end{bmatrix}, \nonumber\\
%%%%%%%%%%%%%%%%%%%%%%%%%%%%%%%%%%%%%%
H_{pd}^{(2)} &=
\begin{bmatrix}
    T_{6} & 0 \\
    0     & 0  \\
    0     & T_{6}
\end{bmatrix}, \hspace{6mm}
%%%%%%%%%%%%%%%%%%%%%%%%%%%%%%%%%%%%%%
H_{dp}^{(2)} =
\begin{bmatrix}
    T_{8} & 0 & 0 \\
    0     & 0 & T_{8}
\end{bmatrix}.
\end{align}
}
Finally, for the $d$-$d$ couplings, we obtain:
%%%%%%%%%%%%%%%%%%%%%%%%%%%%%%%%%%%%%%
{\allowdisplaybreaks
\begin{align}
H^{(1)}_{dd} &=
\begin{bmatrix}
    T_{9}f_{+1}(-\vec{k})  & T_{10}f_{0}(-\vec{k}) & T_{11}f_{-1}(-\vec{k})\\
    T_{10}f_{-1}(-\vec{k}) & T_{12}f_{+1}(-\vec{k}) & T_{10}f_{0}(-\vec{k}) \\
    T_{11}f_{0}(-\vec{k})  & T_{10}f_{-1}(-\vec{k}) & T_{9}f_{+1}(-\vec{k})
\end{bmatrix},\nonumber\\
%%%%%%%%%%%%%%%%%%%%%%%%%%%%%%%%%%%%%%
H^{(2)}_{dd} &=
\begin{bmatrix}
    T_{13}f_{-1}(-\vec{k}) &  T_{14}f_{+1}(-\vec{k}) \\
    T_{15}f_{0}(-\vec{k}) & -T_{15}f_{-1}(-\vec{k})\\
   -T_{14}f_{+1}(-\vec{k})& -T_{13}f_{0} (-\vec{k})
\end{bmatrix},\nonumber\\
%%%%%%%%%%%%%%%%%%%%%%%%%%%%%%%%%%%%%%
H^{(3)}_{dd} &=
\begin{bmatrix}
    T_{13}f_{0}(-\vec{k})  &  T_{15}f_{-1}(-\vec{k}) & -T_{14}f_{+1}(-\vec{k})\\
    T_{14}f_{+1}(-\vec{k}) & -T_{15}f_{0}(-\vec{k})  & -T_{13}f_{-1}(-\vec{k})
\end{bmatrix},\nonumber\\
%%%%%%%%%%%%%%%%%%%%%%%%%%%%%%%%%%%%%%
H^{(4)}_{dd} &=
\begin{bmatrix}
    T_{16}f_{+1}(-\vec{k}) & T_{17}f_{0}(-\vec{k}) \\
    T_{17}f_{-1}(-\vec{k}) & T_{16}f_{+1}(-\vec{k})
\end{bmatrix}
\end{align}}
%%%%%%%%%%%%%%%%%%%%%%%%%%%%%%%%%%%%%%
In the above equations $f$ are $\vec{k}$-dependent functions defined in the previous section, while the amplitudes $T$ are described by the heterostructure geometry and Slater-Koster parameters. 
%%%%%%%%%%%%%%%%% Fig %%%%%%%%%%%%%%%%
\begin{figure}[t]
\includegraphics[width=0.9\linewidth]{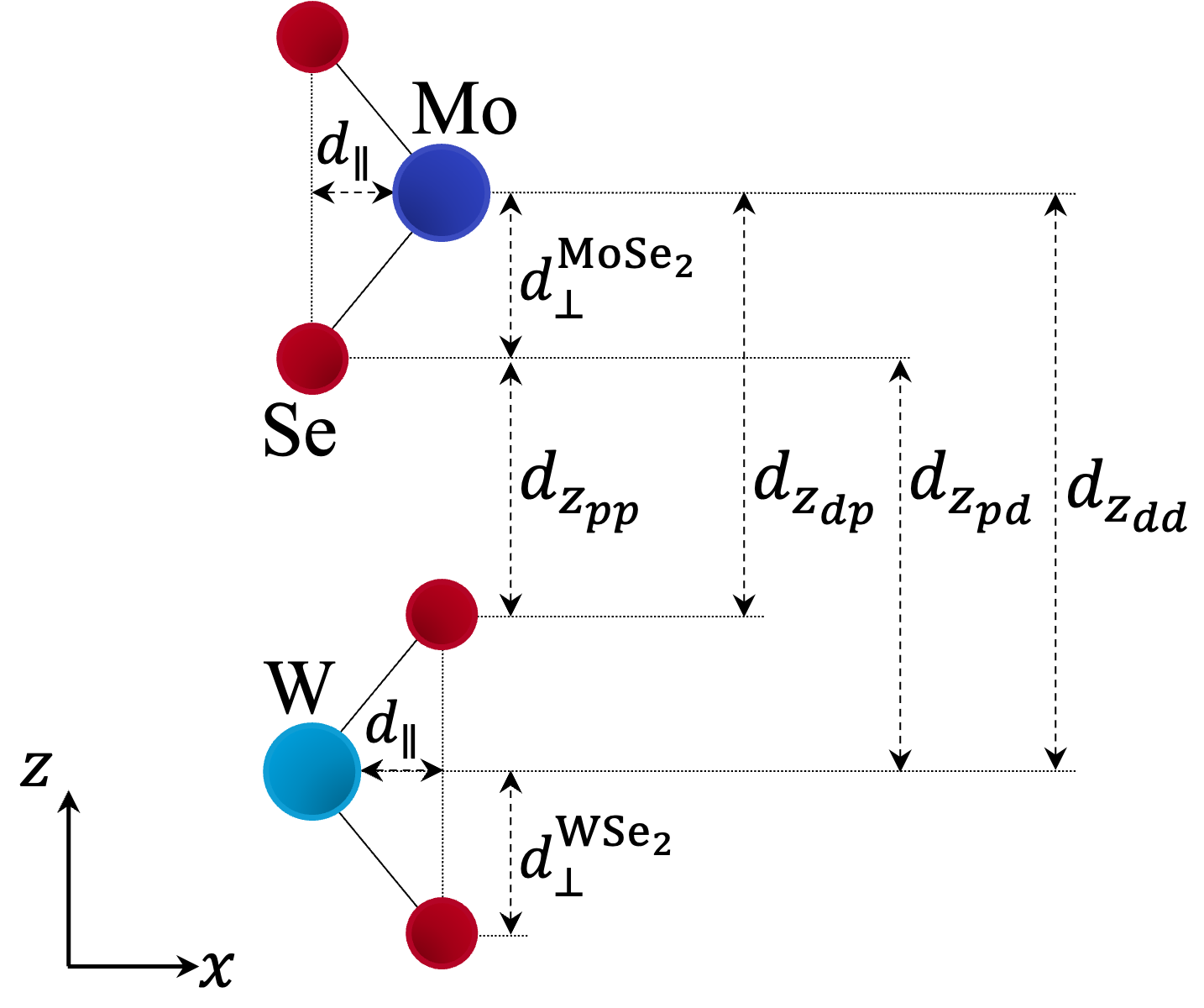}
    \caption{Details of the geometric distances used in the \ac{TB} model construction.}
\label{fig:geoapp}
\end{figure}
%%%%%%%%%%%%%%%%%%%%%%%%%%%%%%%%%%%%%%
These amplitudes parametrize the interlayer Mo-W (subscript $dd$), Se$_2$-Se$_2$ (subscript $pp$), and metal-chalcogen (subscripts $pd$ and $dp$) interactions, respectively. Additional distances on top of those presented in Fig.~\ref{fig:TMDgeometry} are shown in Fig.~\ref{fig:geoapp}. For clarity we introduce also the following distances: $d_{pp}=\sqrt{d_{zpp}^{2}+d_{||}^{2}}$, $d_{dd}=\sqrt{d_{zdd}^{2}+d_{||}^{2}}$, $d_{dp}=d_{zdp}$, $d_{pd}=d_{zpd}$, $d_{zdp}=d_{zdd}-d_{\perp}^{\text{WSe$_2$}}$, $d_{zpd}=d_{zdd}-d_{\perp}^{\text{MoSe$_2$}}$, $d_{zpp}=d_{zdd}-d_{\perp}^{\text{MoSe$_2$}}-d_{\perp}^{\text{WSe$_2$}}$. Numerical values we use are $d_{\perp}=3.323/\sqrt{3}$~\AA, $d_{\perp}^{\text{MoSe$_2$}}=2.869$~\AA , $d_{\perp}^{\text{WSe$_2$}} = 2.880$~\AA , $d_{zdd}=6.40$~\AA. Below we present the formulas for each of the $17$ amplitudes $T$. For the $p$-$p$ couplings the amplitudes are defined as:
%%%%%%%%%%%% T formulas %%%%%%%%%%%%%%
{\allowdisplaybreaks
\begin{align}
    T_{1} &= \frac{1}{4}\left(\left[1-\left(\frac{d_{zpp}}{d_{pp}}\right)^2\right]V_{pp\sigma}+\left[1+\left(\frac{d_{zpp}}{d_{pp}}\right)^2\right]V_{pp\pi}\right), \nonumber\\
    T_{2} &= \frac{1}{2\sqrt{2}}\frac{d_{\|}d_{zpp}}{d_{pp}^2}\left(V_{pp\sigma}-V_{pp\pi}\right), \nonumber\\
    T_{3} &= -\frac{1}{4}\left(\frac{d_{\|}}{d_{pp}}\right)^2\left(V_{pp\sigma}-V_{pp\pi}\right), \nonumber\\
    T_{4} &= -\frac{1}{2}\left(\left(\frac{d_{zpp}}{d_{pp}}\right)^2 V_{pp\sigma}+\left(1-\left(\frac{d_{zpp}}{d_{pp}}\right)^2\right)V_{pp\pi}\right),
\label{eq:TppFormulas}
\end{align}}
Next, for the $p$-$d$ and $d$-$p$ couplings:
%%%%%%%%%%%%%%%%%%%%%%%%%%%%%%%%%%%%%%
{\allowdisplaybreaks
\begin{align}
    T_{5} &= \frac{1}{\sqrt{2}}V_{pd\sigma}, \nonumber\\
    T_{6} &= -\frac{1}{\sqrt{2}}V_{pd\pi},   \nonumber\\
    T_{7} &= \frac{1}{\sqrt{2}}V_{dp\sigma}, \nonumber\\
    T_{8} &= \frac{1}{\sqrt{2}}V_{dp\pi}.
\label{eq:TdpFormulas}
\end{align}
}
%%%%%%%%%%%%%%%%%%%%%%%%%%%%%%%%%%%%%%
Finally, in the case of $d$-$d$ couplings:
\begin{widetext}
{\allowdisplaybreaks
\begin{align}
T_{9} &=
    \frac{1}{8} \left[\left(3-6\left(\frac{d_{zdd}}{d_{dd}}\right)^2+3\left(\frac{d_{zdd}}{d_{dd}}\right)^4\right)V_{dd\sigma} + 
    \left(4-4\left(\frac{d_{zdd}}{d_{dd}}\right)^4\right)V_{dd\pi} +
    \left(1+6\left(\frac{d_{zdd}}{d_{dd}}\right)^2+\left(\frac{d}{d_{dd}}\right)^4\right)V_{dd\delta}\right],\nonumber\\
T_{10} &= 
    \frac{1}{4}\sqrt{\frac{3}{2}}\left(\frac{d_{\|}}{d_{dd}}\right)^{2}
    \left[
    \left(-1+3\left(\frac{d_{zdd}}{d_{dd}}\right)^2\right)V_{dd\sigma}
    - 4\left(\frac{d_{zdd}}{d_{dd}}\right)^2V_{dd\pi} + 
    \left(1+\left(\frac{d_{zdd}}{d_{dd}}\right)^2\right)V_{dd\delta}
    \right],\nonumber\\
T_{11} &= 
    \frac{1}{8} \left(\frac{d_{\|}}{d_{dd}}\right)^{4} \left( 3V_{dd\sigma} - 4V_{dd\pi} + V_{dd\delta} \right),\nonumber\\
T_{12} &= 
    \frac{1}{4} \left[\left(1-6\left(\frac{d_{zdd}}{d_{dd}}\right)^2+9\left(\frac{d_{zdd}}{d_{dd}}\right)^4\right)V_{dd\sigma} +
    12 \left(\left(\frac{d_{zdd}}{d_{dd}}\right)^2-\left(\frac{d_{zdd}}{d_{dd}}\right)^4\right)V_{dd\pi} +
     \left(3-6\left(\frac{d_{zdd}}{d_{dd}}\right)^2+3\left(\frac{d_{zdd}}{d_{dd}}\right)^4\right)V_{dd\delta}\right],\nonumber\\
T_{13} &=  
    \frac{1}{4}\frac{d_{\|}}{d_{dd}}\frac{d_{zdd}}{d_{dd}}
    \left[
    \left(3-3\left(\frac{d_{zdd}}{d_{dd}}\right)^2\right)V_{dd\sigma} +
    4\left(\frac{d_{zdd}}{d_{dd}}\right)^2V_{dd\pi} -
    \left(3+\left(\frac{d_{zdd}}{d_{dd}}\right)^2\right)V_{dd\delta}
    \right],\nonumber\\
T_{14} &= 
    -\frac{1}{4} \left(\frac{d_{\|}}{d_{dd}}\right)^{3}\frac{d_{zdd}}{d_{dd}} \left( 3V_{dd\sigma} - 4V_{dd\pi} + V_{dd\delta} \right),\nonumber\\
T_{15} &= 
    \frac{1}{2}\sqrt{\frac{3}{2}}\frac{d_{\|}}{d_{dd}}\frac{d_{zdd}}{d_{dd}}
    \left[
    \left(-1+3\left(\frac{d_{zdd}}{d_{dd}}\right)^2\right)V_{dd\sigma} +
    \left(2-4\left(\frac{d_{zdd}}{d_{dd}}\right)^2\right)V_{dd\pi} +  
    \left(-1+\left(\frac{d_{zdd}}{d_{dd}}\right)^2\right)V_{dd\delta}
    \right],\nonumber\\
T_{16} &=  
    \frac{1}{2}\left[ 3\left(\left(\frac{d_{zdd}}{d_{dd}}\right)^2-\left(\frac{d_{zdd}}{d_{dd}}\right)^4\right)V_{dd\sigma} + 
    \left(1-3\left(\frac{d_{zdd}}{d_{dd}}\right)^2+4\left(\frac{d_{zdd}}{d_{dd}}\right)^4\right)V_{dd\pi} +
    \left(1-\left(\frac{d_{zdd}}{d_{dd}}\right)^4\right)V_{dd\delta}\right],\nonumber\\
T_{17} &= 
    -\frac{1}{2}\left(\frac{d_{\|}}{d_{dd}}\right)^{2}
    \left[
    3\left(\frac{d_{zdd}}{d_{dd}}\right)^2 V_{dd\sigma} +
    \left(1-4\left(\frac{d_{zdd}}{d_{dd}}\right)^2\right) V_{dd\pi} + 
    \left(-1+\left(\frac{d_{zdd}}{d_{dd}}\right)^2\right) V_{dd\delta}
    \right],
\label{eq:TddFormulas}
\end{align}
}
\end{widetext}
%%%%%%%%%%%%%%%%%%%%%%%%%%%%%%%%%%%%%%

The parameters of the full \ac{TB} model for the MoSe$_2$/WSe$_2$ heterostructure are summarized in the Tab.~\ref{tab:SKParametersFull}. Furthermore, in Tab.~\ref{tab:OrbContributionTB} we present the orbital contributions to the energy bands for the even subspace, presented for the atomic orbitals with the main contribution to the top \ac{VB} and bottom \ac{CB}. 

%%%%%%%%%%%%%%%%% Tab %%%%%%%%%%%%%%%%
\begin{table}[t]
\centering 
\caption{Slater-Koster parameters for the MoSe$_2$/WSe$_2$ heterostructure in the full/simplified \ac{TB} model.}
\vspace{2mm}
\setlength\cellspacetoplimit{2pt}
\setlength\cellspacebottomlimit{4pt}
\begin{tabular}{ | O{P{1.0cm}} | O{P{2.3cm}} | O{P{2.3cm}} | O{P{2.3cm}} | } \hline\hline
                    & MoSe$_2$ & WSe$_2$ & interlayer \\\hline\hline
    $E_d$                   & $-0.140$/$-0.240$ & $-0.220$/$-0.210$ &  - \\
    $E_{p_1}$               & $-5.060$/$-5.258$ & $-4.234$/$-4.419$ &  - \\
    $E_{p_0}$               & $-5.268$/$-5.620$ & $-4.620$/$-4.890$ &  - \\
    $V_{dp\sigma}$          & $-2.980$/$-2.976$ & $-3.251$/$-3.262$ &  $-0.044$/ - \\
    $V_{pd\sigma}$          & -                & -                &  $-1.344$/ - \\
    $V_{dp\pi}$             & $ 1.073$/$ 1.175$ & $ 1.060$/$ 1.020$ &  $ 0.411$/ - \\
    $V_{pd\pi}$             & -                & -                &  $-1.780$/ - \\
    $V_{dd\sigma}$          & $-1.028$/$-0.923$ & $-1.240$/$-1.223$ &  $ 0.196$/$-0.500$ \\
    $V_{dd\pi}$             & $ 0.838$/$ 0.751$ & $ 0.817$/$ 0.860$ &  $-0.480$/$ 2.000$ \\
    $V_{dd\delta}$          & $ 0.229$/$ 0.225$ & $ 0.255$/$ 0.238$ &  $ 0.091$/$-0.330$ \\
    $V_{pp\sigma}$          & $ 1.490$/$ 1.397$ & $ 1.258$/$ 1.079$ &  $ 1.393$/$-1.101$ \\
    $V_{pp\pi}$             & $-0.550$/$-0.468$ & $-0.350$/$-0.366$ &  $-0.012$/$-0.155$ \\
    $E_{p_1}^\text{odd}$    & $-5.051$/$-5.398$ & $-4.132$/$-4.559$ &  - \\
    $E_{p_0}^\text{odd}$    & $-5.236$/$-5.598$ & $-4.554$/$-5.029$ &  - \\
    $E_d^\text{odd}$        & $-0.112$/$-0.380$ & $-0.032$/$-0.350$ &  - \\
		$\lambda_{M}$           & $ 0.093$/$ 0.093$   & $ 0.236$/$ 0.236$ &  - \\
		$\lambda_{X_{2}}$       & $ 0.175$/$ 0.100$   & $-0.275$/$-0.195$ &  - \\\hline\hline
\end{tabular}
\label{tab:SKParametersFull}
\end{table}
%%%%%%%%%%%%%%%%%%%%%%%%%%%%%%%%%%%%%%

%%%%%%%%%%%%%%% Tab %%%%%%%%%%%%%%%%%%
\begin{table}[t]
\centering 
\caption{Orbital contribution from full \ac{TB} model for the even subspace presented for the atomic orbitals with the main contribution to the top \ac{VB} and bottom \ac{CB}. Numbers are presented in \% for a given high symmetry point.}
\vspace{2mm}
\setlength\cellspacetoplimit{2pt}
\setlength\cellspacebottomlimit{4pt}
\begin{tabular}{ | O{P{1.2cm}} O{P{1.2cm}} | O{P{1.35cm}} O{P{1.35cm}} O{P{1.35cm}} O{P{1.35cm}} | } \hline\hline

                             &           & K, \ac{VB}   & K, \ac{CB} & Q, \ac{CB} & $\Gamma$, \ac{VB}  \\ 
     & & $\downarrow$/$\uparrow$ & $\downarrow$/$\uparrow$ & $\downarrow$/$\uparrow$ & $\downarrow$/$\uparrow$ \\ \hline\hline
                                 & $4p_{-1}$ & -        & 0\%/12\% & 2\%/0\%  & -         \\
    Se$_2^\text{top}$    & $4p_{0}$  & -        & -        & 7\%/1\%  & 3\%/3\%   \\ 
                                 & $4p_{+1}$ & -  & -        & 1\%/0\%  & -         \\\hline
                                 & $4d_{-2}$ & -        & - & 13\%/1\% & -   \\
    Mo                           & $4d_{0}$  & -        & 0\%/83\% & 11\%/1\%  & 32\%/32\%  \\
                                 & $4d_{+2}$ & -  & -        & 3\%/0\%  & -   \\\hline\hline
                                 & $4p_{-1}$ & - & -        & 4\%/0\%  & -         \\
    Se$_2^\text{bottom}$ & $4p_{0}$  & -        & -        & 9\%/0\%  & -   \\
                                 & $4p_{+1}$ & -        & -        & 1\%/0\%  & -         \\\hline
                                 & $5d_{-2}$ & 97\%/0\% & -        & 1\%/0\%  & -         \\
    W                            & $5d_{0}$  & -        & 1\%/0\%  & 17\%/1\%  & 15\%/15\% \\
                                 & $5d_{+2}$ & -        & 0\%/1\%  & 24\%/0\% & -  \\\hline\hline

\end{tabular}
\label{tab:OrbContributionTB}
\end{table}
%%%%%%%%%%%%%%%%%%%%%%%%%%%%%%%%%%%%%%

%%%%%%%%%%%%%%%%%%%%%%%%%%%%%%%%%%%%%%%% Appendix D %%%%%%%%%%%%%%%%%%%%%%%%%%%%%%%%%%%%%%
\section{Basis transformation for electric field}
\label{section:AppendixEfield}
%%%%%%%%%%%%%%%%%%%%%%%%%%%%%%%%%%%%%%

In Section~\ref{section:Efield} we have introduced the electric field Hamiltonian for an applied homogeneous perpendicular electric field $E_z$. Here we present the details of the Hamiltonian transformation to the dimer basis for a monolayer electric field Hamiltonian block.

Starting with the single-atomic basis (each Se atom taken separately), where the Hamiltonian for a single layer of \ac{TMDC} crystals can be defined more intuitively in the basis: $\{d_{-2},d_{0},d_{2},p_{-1}^{t},p_{0}^{t},p_{1}^{t},d_{-1},d_{1},p_{-1}^{b},p_{0}^{b},p_{1}^{b}\}$, where $p^{t}$ and $p^{b}$ correspond to $p$-orbitals on top and bottom Se atoms. The electric field Hamiltonian can be written explicitly for a given spin as an $11\times11$ matrix containing diagonal terms only $H_E^\mathrm{atom}=\textrm{diag}(V_{M},V_{M},V_{M},V_{X^t},V_{X^t},V_{X^t},V_{M},V_{M},V_{X^b},V_{X^b},V_{X^b})$. To reformulate this Hamiltonian to form given in main text for basis using dimer orbitals, we construct the unitary transformation matrix between the both bases:
%%%%%%%%%%%%%% Eq Efield %%%%%%%%%%%%%
\begin{equation}
\hat{U}_E = \frac{1}{\sqrt{2}}\begin{bmatrix*}
    \sqrt{2} & 0        & 0        & 0 & 0 & 0 & 0        & 0        & 0 & 0 & 0 \\
    0        & \sqrt{2} & 0        & 0 & 0 & 0 & 0        & 0        & 0 & 0 & 0 \\
    0        & 0        & \sqrt{2} & 0 & 0 & 0 & 0        & 0        & 0 & 0 & 0 \\
    0        & 0        & 0        & 1 & 0 & 0 & 0        & 0        & 1 & 0 & 0 \\
    0        & 0        & 0        & 0 & 1 & 0 & 0        & 0        & 0 &-1 & 0 \\
    0        & 0        & 0        & 0 & 0 & 1 & 0        & 0        & 0 & 0 & 1 \\
    0        & 0        & 0        & 0 & 0 & 0 & \sqrt{2} & 0        & 0 & 0 & 0 \\
    0        & 0        & 0        & 0 & 0 & 0 & 0        & \sqrt{2} & 0 & 0 & 0 \\
    0        & 0        & 0        & 1 & 0 & 0 & 0        & 0        &-1 & 0 & 0 \\
    0        & 0        & 0        & 0 & 1 & 0 & 0        & 0        & 0 & 1 & 0 \\
    0        & 0        & 0        & 0 & 0 & 1 & 0        & 0        & 0 & 0 &-1
\end{bmatrix*}.
\end{equation}
%%%%%%%%%%%%%%%%%%%%%%%%%%%%%%%%%%%%%%
By acting the transformation matrix on the $H_E^\mathrm{atom}$ Hamiltonian, $\hat{U}_{E}\hat{H}_E^{\text{atom}}\hat{U}^{-1}_{E}$, we obtain the electric field Hamiltonian given by Eq.~(\ref{eq:EfieldInclusionGeneral}).

%%%%%%%%%%%%%%%%%%%%%%%%%%%%%%%%%%%%%%%% Appendix E %%%%%%%%%%%%%%%%%%%%%%%%%%%%%%%%%%%%%%
%%%%%%%%%%%%%%%%%%%%%%%%%%%%%%%%%%%%%%
\section{Simplified tight-binding model}
\label{section:AppendixTBsimple}
%%%%%%%%%%%%%%%%%%%%%%%%%%%%%%%%%%%%%%

%%%%%%%%%%%%%%%%% Fig %%%%%%%%%%%%%%%%
\begin{figure}[t]
\includegraphics[width=0.9\linewidth]{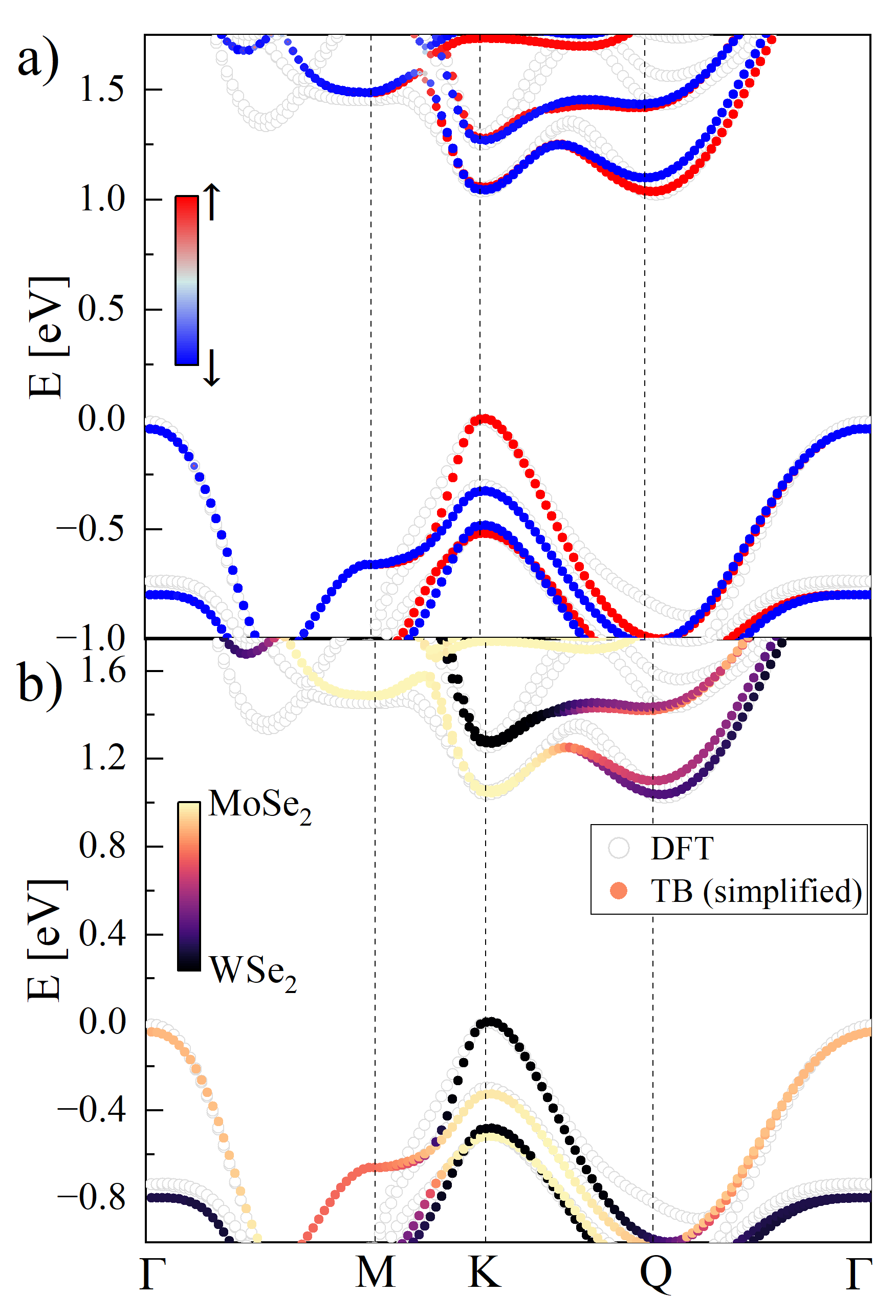}
    \caption{Electronic structure of the MoSe$_2$/WSe$_2$ for the simplified \ac{TB} model (full dots), presented in a comparison with \ac{DFT} (open circles). The energy spectrum is presented on the $\Gamma$-$M$-$K$-$Q$-$\Gamma$ path. The panels show (a) spin- and (b) layer-resolved electronic structures. Both spin and layer are decoded by color, where red/blue denotes spin up/down, while yellow/black denotes MoSe$_2$/WSe$_2$ layer, respectively.}
\label{fig:TBsimple}
\end{figure}
%%%%%%%%%%%%%%%%%%%%%%%%%%%%%%%%%%%%%%

Having the \ac{TB} model taking into account all interlayer orbital interactions, we present a simplified approach reproducing main MoSe$_2$/WSe$_2$ heterostructure features known from \ac{DFT}. Guided by the majority orbital contributions to the \ac{VB} and \ac{CB} at high symmetry points, we simplify the interlayer interactions Hamiltonian block $H_{\text{inter}}$, defined in Eq.~(\ref{eq:InterHamiltonianFull}), by taking into account only the $p_0$-$p_0$, $d_0$-$d_0$, $d_{-2}$-$d_{+2}$ and $d_{+2}$-$d_{-2}$ couplings from the even orbital subspace. This reduces the $H_{\text{inter}}$ to the following form:
%%%%%%%%%%%%% Eq H inter %%%%%%%%%%%%%
\begin{equation}
    H_{\text{inter}} =
    \begin{bmatrix}
        H_{dd}^\text{ev-ev} &0 &0 & 0 \\
       0 & H_{pp}^\text{ev-ev} & 0 & 0\\
				0 & 0 & 0 & 0 \\
        0 & 0 & 0 & 0
    \end{bmatrix},
\label{eq:InterHamiltonianSimple}
\end{equation}
%%%%%%%%%%%%%%%%%%%%%%%%%%%%%%%%%%%%%%
where the coupling blocks $H_{dd}^\text{ev-ev}$ and $H_{pp}^\text{ev-ev}$ are simplified to
%%%%%%%%%%% Eq H inter dd %%%%%%%%%%%%
\begin{equation}
\begin{aligned}
&H_{dd}^\text{ev-ev} = \\&=
\begin{bmatrix}
    0 & 0 & T_{11}f_{-1}(-\vec{k})\\
    0 & T_{12}f_{+1}(-\vec{k}) & 0 \\
    T_{11}f_{0}(-\vec{k}) & 0 & 0
\end{bmatrix},
\end{aligned}
\label{eq:InterddHamiltonian}
\end{equation}
%%%%%%%%%%%%%%%%%%%%%%%%%%%%%%%%%%%%%%
and
%%%%%%%%%%% Eq H inter pp %%%%%%%%%%%%
\begin{equation}
    H_{pp}^\text{ev-ev} =
    \begin{bmatrix}
        0 & 0 & 0 \\
        0 & T_{4}f_{+1}(\vec{k}) & 0 \\
        0 & 0 & 0
    \end{bmatrix} 
\label{eq:InterppHamiltonian}
\end{equation}
%%%%%%%%%%%%%%%%%%%%%%%%%%%%%%%%%%%%%%
Functions $f$ and amplitudes $T$ have been defined in the previous section.

This gives the 5 interlayer Slater-Koster parameters (compared to 9 for the full \ac{TB} model) and 26 intralayer parameters (13 for each layer), creating the subspace of 31 Slater-Koster parameters that have to be parametrized. In the Tab.~\ref{tab:SKParametersFull} we present full and simplified \ac{TB} model parameters. We keep the parameters for both even and odd blocks of monolayer Hamiltonians, however good results can be obtained in even-only subspace. 

Fig.~\ref{fig:TBsimple}(a,b) presents the electronic structure obtained by diagonalizing the Hamiltonian with the simplified interlayer interactions. The type-II band alignment has been captured, with the direct $K$-$K$ energy gap of the order of $1$ eV. The dispersion is characterised by the degenerate \ac{CB} minima in $K$ and $Q$ points, as well as the degenerate \ac{VB} maxima in $K$ and $\Gamma$ points, and stays in good agreement with the \ac{DFT} results. The effective masses around $K$ valley are also consistent with \ac{DFT}. The effect of spin-valley locking in close vicinity of the $K$ valley has been obtained. The intralayer spin splittings due to the \ac{SOC} in both \ac{CB} and \ac{VB} in $K$ point are established to be for MoSe$_2$ $\Delta_{\text{SOC}}^{CB}=12$~meV and $\Delta_{\text{SOC}}^{VB}=189$~meV, while for WSe$_2$ $\Delta_{\text{SOC}}^{CB}=30$~meV and $\Delta_{\text{SOC}}^{VB}=521$~meV. However, the effects of spin mixing due to the interactions between even and odd orbitals have not been captured for the low-energy bands. Fig.~\ref{fig:TBsimple}(b) shows that the electron in $Q$ valley is delocalized between distinct layers, nevertheless the delocalization is not captured quantitatively. Subsequently, we have performed the analysis of orbital contributions. We note that they stay in quantitative agreement with the full \ac{TB} model results. This is understood from the fact that the low-energy bands are mainly composed of the even orbitals. 

%%%%%%%%%%%%%%%%% Fig %%%%%%%%%%%%%%%%
\begin{figure}[t]
\includegraphics[width=\linewidth]{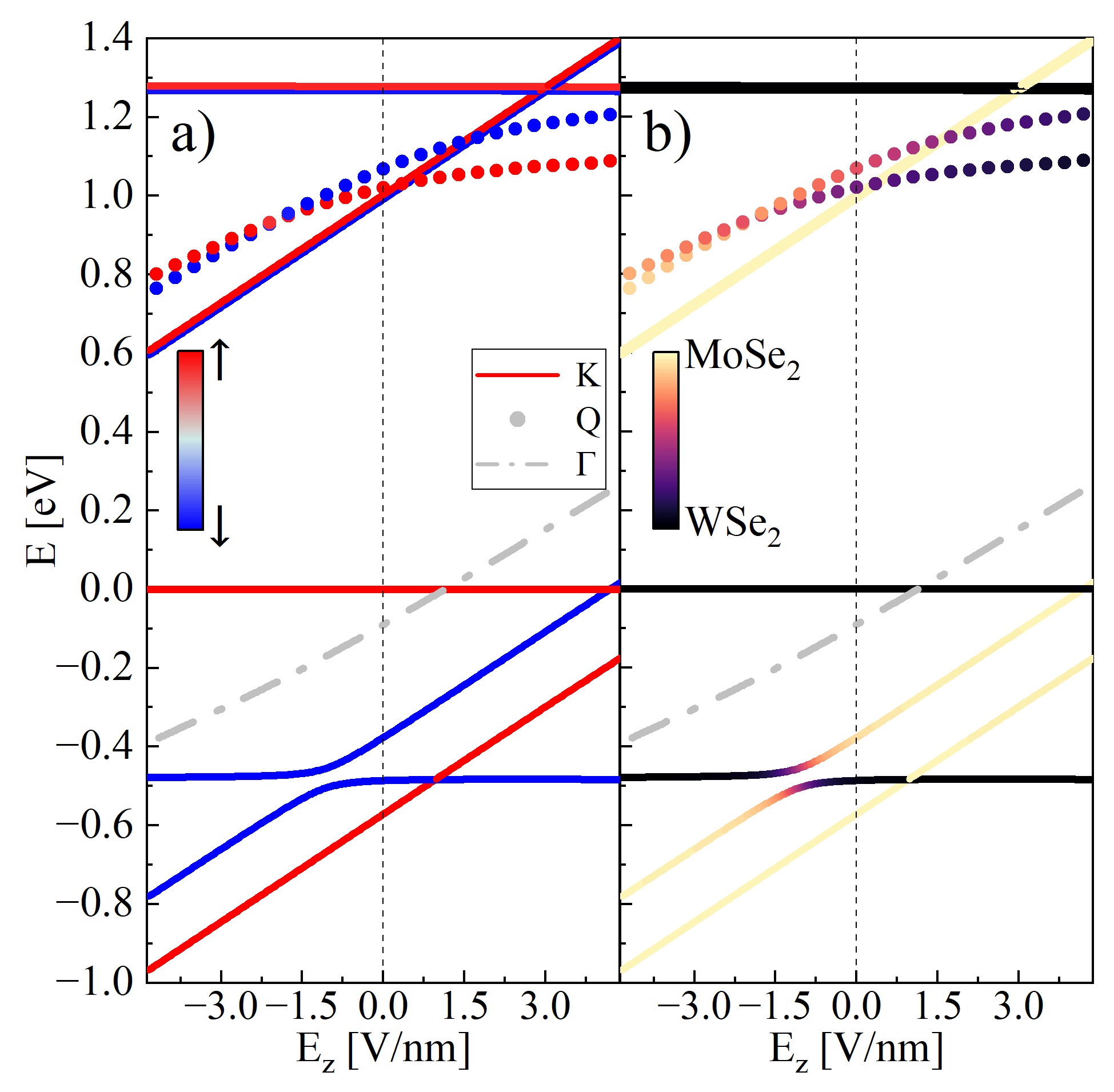}
    \caption{{The effect of a vertical electric field on the electronic structure in the simplified \ac{TB} model.} Energies of selected bands are presented as a function of the applied $E_z$ for $K$, $Q$ and $\Gamma$. The top of the \ac{VB} in $K$ is set to $E=0$~eV. Different type of lines represent respectively $K$ -- continues line, $Q$ -- dashed line, $\Gamma$ -- dashed-dotted, as denoted on the scheme. colors correspond to (a) spin, and (b) layer composition, respectively. The color schemes are denoted on the plots. Due to the degeneracy of \ac{VB} in the $\Gamma$ point, the spin/layer-resolved notation was not introduced.}
\label{fig:ElectricFieldTBsimple}
\end{figure}
%%%%%%%%%%%%%%%%%%%%%%%%%%%%%%%%%%%%%%

In the next step, following the methodology described in Section~\ref{section:Efield} we perform a study of the electronic structure evolution as a function of applied electric field $E_z$. The spin and layer contributions to the high symmetry points $K$, $Q$, and $\Gamma$ for  bands around the energy gap are presented on Fig.~\ref{fig:ElectricFieldTBsimple}. We note that, unlike in the case of the full \ac{TB} model, the applied electric field does not change the spin localization for both $K$ and $Q$ points. The \ac{CB}s \ac{SOC} splittings in $K$ point are not affected by the applied $E_z$, while in the \ac{VB} the splitting increase with the applied negative electric field, in agreement with the full \ac{TB} model. As observed within \ac{DFT} and full \ac{TB} approximation, the spin-splitting between the two lowest \ac{CB}s in $Q$ point increases with the positive $E_z$ and decreases with the negative $E_z$, respectively. Furthermore, the crossing of the \ac{CB}s in the $Q$ point as the negative applied $E_z$ increases has been observed, changing the character of the $K$-$Q$ energy gap. However, unlike in the case of full \ac{TB} model, the effect of spin mixing around the crossing points have not been obtained. 

In Fig.~\ref{fig:ElectricFieldTBsimple}(b) we present the layer contribution. We note that the vertical electric field affects the coupling between layers, in agreement with the full \ac{TB} model results. However, we point out that the effect of switching the leading layer contribution obtained withing the simplified \ac{TB} model has been captured not only in the $Q$ point for the \ac{CB}s, but also in the $K$ point for the \ac{VB}s, which stays in agreement with the \ac{DFT} results. The intralayer energy gap for WSe$_2$ remains almost constant when applying the electric field, however the intralayer MoSe$_2$ energy gap changes due to the effects of layer mixing in the \ac{VB}, which has not been observed in the full \ac{TB} model. The direct interlayer energy gap decreases when applying $E_z$. Overall, \ac{DFT}, full and simplified \ac{TB} results agree on the character of the energy bands modulation with the applied electric field.

%%%%%%%%%%%%%%%%%%%%%%%%%%%%%%%%%%%%%%% Bibliography %%%%%%%%%%%%%%%%%%%%%%%%%%%%%%%%%%%%%

\newpage*
%\bibliographystyle{apsrev4-2}
%\bibliography{Bibliography}

%apsrev4-2.bst 2019-01-14 (MD) hand-edited version of apsrev4-1.bst
%Control: key (0)
%Control: author (72) initials jnrlst
%Control: editor formatted (1) identically to author
%Control: production of article title (-1) disabled
%Control: page (0) single
%Control: year (1) truncated
%Control: production of eprint (0) enabled
%

\end{document}